\documentclass[prapplied,aps,two column,reprint,superscriptaddress,longbibliography]{revtex4-2}
\usepackage{bm}
\usepackage[colorlinks=true,linkcolor=blue,citecolor=blue]{hyperref}
\usepackage{times}
\usepackage{amsmath}
\usepackage{amssymb}
\usepackage{amsthm}
\usepackage{amsfonts}
\usepackage{enumerate}
\usepackage{latexsym}
\usepackage{ifpdf}
\newcommand{\beq}{\begin{equation}}
\newcommand{\eeq}{\end{equation}}
\usepackage{graphicx}
\usepackage{makeidx}
\usepackage{color}
\usepackage{physics}
\usepackage{siunitx}

\begin{document}
\title{Electron trapping and detrapping in an oxide two-dimensional electron gas: The role of ferroelastic twin walls}

\author {Shashank Kumar Ojha}
	\email{shashank@iisc.ac.in}
	\affiliation  {Department of Physics, Indian Institute of Science, Bengaluru 560012, India}
	\author {Sankalpa Hazra}
	\affiliation  {Department of Physics, Indian Institute of Science, Bengaluru 560012, India}
	\author{Prithwijit Mandal}
	\affiliation  {Department of Physics, Indian Institute of Science, Bengaluru 560012, India}
        \author{Ranjan Kumar Patel}
	\affiliation  {Department of Physics, Indian Institute of Science, Bengaluru 560012, India}
	\author{Shivam Nigam}
	\affiliation  {Department of Physics, Indian Institute of Science, Bengaluru 560012, India}
	\author {Siddharth Kumar}
	\affiliation  {Department of Physics, Indian Institute of Science, Bengaluru 560012, India}
        \author {S. Middey}
	\email{smiddey@iisc.ac.in}
	\affiliation  {Department of Physics, Indian Institute of Science, Bengaluru 560012, India}

\setcounter{dbltopnumber}{2}

\begin{abstract}

The choice of electrostatic gating over the conventional chemical doping for phase engineering of quantum materials is attributed to the fact that the former can reversibly tune the carrier density without affecting the system's level of disorder. However, this proposition seems to break down in field-effect transistors involving SrTiO$_3$ (STO) based two-dimensional electron gases. Such peculiar behavior is associated with the electron trapping under an external electric field. However, the microscopic nature of trapping centers remains an open question. In this paper, we investigate electric field-induced charge trapping/detrapping phenomena at the conducting interface between band insulators $\gamma$-Al$_2$O$_3$ and STO. Our transport measurements reveal that the charge trapping under +ve back gate voltage ($V_g$) above the tetragonal to cubic structural transition temperature ($T_c$) of STO is contributed by the electric field-assisted thermal escape of electrons from the quantum well,  and the clustering of oxygen vacancies (OVs) as well. We observe an additional source of trapping below the $T_c$, which arises from the trapping of free carriers at the ferroelastic twin walls of STO.  Application of -ve $V_g$ results in a charge detrapping, which vanishes above $T_c$ also. This feature demonstrates the crucial role of structural domain walls in the electrical transport properties of STO based heterostructures. The number of trapped (detrapped) charges at (from) the twin wall is controlled by the net polarity of the wall and is completely reversible with the sweep of $V_g$.
  \end{abstract}

\maketitle

\section{Introduction}
\begin{figure*}
	\centering{
		{~}\hspace*{-0cm}
		\includegraphics[scale=0.55]{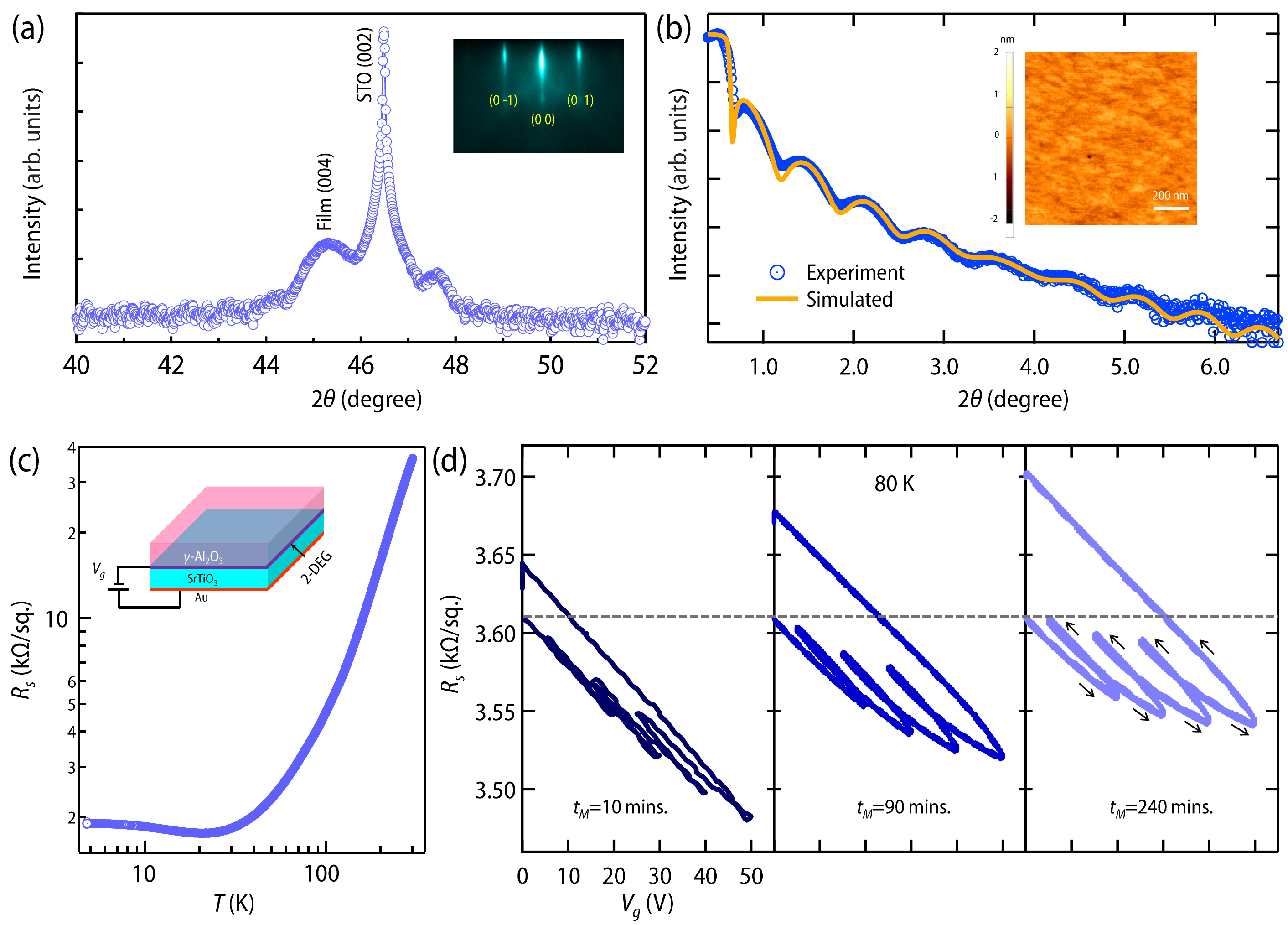}
		\caption{ (a) XRD pattern of GAO/STO heterostructure around the STO (0 0 2) peak. Right inset is  RHEED image of the GAO film taken after cooling down to room temperature.  (b) X-ray reflectivity (XRR) pattern of GAO film. Inset shows  AFM image of GAO film.  (c) Temperature dependence of the $R_S$ of the GAO/STO heterostructure. Inset shows device geometry for resistance measurement of 2-DEG under back gating configuration. (d) Variation of $R_{s}$ at 80 K when $V_g$ is swept  from 0 V $\rightarrow$ 20 V  $\rightarrow$  5 V  $\rightarrow$  30 V  $\rightarrow$  15 V  $\rightarrow$  40 V  $\rightarrow$  25 V  $\rightarrow$  50 V  $\rightarrow$  0 V. Total measurement time $t_M$ for this entire voltage sweep: 10 minutes (left panel), 90 minutes (middle panel), 240 minutes (right panel).  } \label{fig:1}}
\end{figure*}

The field-effect transistor (FET), based on the principle of electrostatic gating of  semiconductor-based heterostructures is the foundation of  modern electronics.   In order to achieve higher transistor packing density beyond the scaling limitations of silicon-based FETs,  several alternative classes of materials  are being explored~\cite{Ahn:2003p1015,Yandenberghe:2017p1,collins:2018p390}.   The quasi two-dimensional electron gas (2-DEG), confined at the surface or at the interface of complex oxides is one such promising platform~\cite{ohtomo:2004p423,Mannhart:2010p1607}.  These oxide based 2-DEG systems  host several unique electronic and magnetic phenomena such as ferromagnetism, multiband superconductivity, spin-orbital texture, topological Hall effect etc.~\cite{mannhart:2008p1027,lee:2013p703,brinkman:2007p493,reyren:2007p1196,stemmer:2014p151,king:2014p1,vaz:2019p1187,Ojha:2020p2000021}, which can further be tuned by the application of an electric field~\cite{thiel:2006p1942,caviglia:2008p624,caviglia:2010p126803,bi:2014p1}.  Owing to the very large dielectric constant of SrTiO$_3$ (STO)~\cite{Weaver:1959p274,Neville:1972p2124,Viana:1994p601}, the electric field control of STO based 2-DEG heterostructures can be  achieved through back gating configuration (see Fig. \ref{fig:1})~\cite{thiel:2006p1942,caviglia:2008p624,Christensen:2016p021602,biscaras:2014p6788,liu:2015p062805,yin:2020p017702}. In this geometry, the conducting interface itself acts as one plate of the capacitor and  an electrode attached to the bottom of the substrate acts as the other plate. Thus,  it is expected that the carrier density of the 2-DEG channel  would be reversibly tuned by varying back-gate voltage ($V_{g}$), as the application of  positive  (negative) $V_{g}$  dopes electron (hole) in the 2-DEG channel.  Surprisingly,   an irreversibility in the sheet resistance ($R_S$) was observed in the case of several STO based 2-DEG systems when the $V_{g}$ was repeatedly swept between zero and +ve $V_{g}$ ~\cite{biscaras:2014p6788,liu:2015p062805,yin:2020p017702,Bal:2017p081604}.  Such irreversibility in $R_S$ appears due to the  loss of conduction electrons from the 2-DEG under +ve $V_g$, leading to the gradual increase of $R_S$ with time under a fixed +ve $V_g$.   However, there is no consensus about the microscopic origin behind such a peculiar behavior,  and  mechanisms such as the thermal escape of electrons from the quantum well under electric field~\cite{biscaras:2014p6788} and trapping of  carriers in the mid-gap states formed due to electric field induced clustering of oxygen vacancies (OVs)~\cite{yin:2020p017702} have been proposed.

 While the number of trapped charges for the above-mentioned processes would depend on the level of structural  defects present in  the STO substrate and defects (such as OV) created at the film/substrate interface during the growth~\cite{Chang:2014p3522,Gunkel:2020p120505}, there can be another inherent source of charge trapping in  STO.   It is well established that STO undergoes  a cubic to tetragonal ferroelastic phase transition around $T_c$ = 105 K~\cite{Cowley:1964pA981}, leading to a dense network of twin walls along certain crystallographic directions.
 Presence of intrinsic strain gradient along with rotopolar and trilinear coupling in the tetragonal phase further makes these twin walls  intrinsically polar~\cite{Schiaffino:2017p137601,Frenkel:2017p1203,Schiaffino:2017p137601,Salje:2013p247603,Zubko:2007p167601}. These polar twin walls  can trap a significant amount of conduction electrons from the 2DEG~\cite{Kalisky:2013p1091,Honig:2013p1112}.
 Owing to its intrinsic polarity, the electric field would be an effective tool to further tune the static polarization of such  twin walls, embedded in the nonpolar matrix of STO~\cite{Ma:2016p257601}, and may have a dramatic impact on  macroscopic electrical transport behavior. However,  signature of such nanoscale charge trapping centers  in  electrostatic gating experiments have not been demonstrated so far.

In this work, we have focussed on the electric field induced charge trapping phenomena in  $\gamma$-Al$_2$O$_3$/SrTiO$_3$ (GAO/STO) heterostructure around the  ferroelastic transition of STO.  GAO has a cubic spinel structure with lattice constant 7.911 \AA,   almost twice the lattice constant of STO (3.905 \AA). Excellent match between oxygen sublattices  of these compounds along [0 0 1] facilitates high-quality epitaxial growth of GAO on STO substrate, and GAO/STO heterostructure exhibits the highest mobility among all STO based 2DEGs realized without modulation doping~\cite{chen:2013p16}.     In contrast to the polar catastrophe driven 2-DEG in LaAlO$_3$/STO (LAO/STO) interface, the conducting interface in GAO/STO emerges due to creation of OVs in the STO side. The presence of excess OVs in GAO/STO~\cite{schutz:2017p161409,christensen:2017p1700026} makes it more susceptible to charge trapping under  back gating experiments and needs special attention.  We found that  the charge trapping in this heterostructure   can not be accounted solely by the  existing proposed mechanisms for the LAO/STO heterostructures. Apart from the loss of carriers due to both thermal escape and  OV clustering under the application of +ve $V_g$, there is an additional component. Most importantly,  this contribution  vanishes above 100 K and increases with the increase of applied field.  Application of negative $V_g$ results in charge detrapping, which also disappears above 100 K.  We attribute this newly observed charge trapping/detrapping feature of carriers to the ferroelastic twin walls of STO.

\section{Experimental details}
15 unit cell thick GAO film was grown on as received mixed terminated single crystalline STO (0 0 1) substrate (5  $\times$ 5 $\times$ 0.5 mm$^3$) by a pulsed laser deposition (PLD) system. The growth was monitored by in-situ high pressure RHEED (reflection high energy electron diffraction). The thin film deposition was carried out at 500 $^\circ$C and the vacuum of the PLD chamber was around 10$^{-6}$ Torr.  A KrF excimer laser was used for the film deposition.  Intense streaks of specular (0,0) and off-specular (0,-1) and (0,1) reflections in the RHEED image [inset of Fig.~\ref{fig:1}(a)] confirm smooth surface morphology and excellent crystallinity of the film.  A  Rigaku Smartlab X-ray diffractometer was used to record  X-ray reflectivity (XRR) and X-ray diffraction (XRD) patterns using Cu $K_{\alpha}$ radiation. Fig.~\ref{fig:1}(a) shows a 2$\theta$-$\omega$ XRD  scan of GAO/STO heterostructure.  As expected for an epitaxial growth on STO (001) substrate,  the pattern consists of a broad GAO film peak,  STO substrate peak along with thickness fringes~\cite{Cao:2016p1}.  The out of plane lattice constant is found to be 8.01 \AA. Atomic force microscopy (AFM) imaging (inset of Fig.~\ref{fig:1}b) further confirms  excellent surface smoothness (mean squared roughness $\simeq$ 180 pm) of the film. The fitting of X-ray reflectivity data [Fig.~\ref{fig:1}(b)] using Genx program~\cite{Genxprogram} finds the thickness of the film to be 120.45 \AA, which is close to the expected value for a 15 unit cell  thick film.

100 nm gold was sputtered at the back side of STO substrate for back gate experiments [Fig.~\ref{fig:1}(c)]. An ultrasonic wire bonder was used to make ohmic contacts with the interface in van der Pauw geometry. The sheet resistance was measured in van der Pauw geometry using ${dc}$ delta mode with  a Keithley 6221 current source and a Keithley 2182A nano-voltmeter. The gate voltage was applied by a Keithley 2450 sourcemeter. The leakage current is at most 2 nA for all the gating experiments performed in this article. All the electrical measurement meters, along with the temperature controller had a common ground connection.   Temperature dependent electrical resistance measurement confirms metallic interface [Fig.~\ref{fig:1}(c)] between GAO and STO, similar to previous reports~\cite{chen:2013p16,Cao:2016p1}.

\section{Results and Discussions }

\begin{figure*}
	\centering{
		{~}\hspace*{-0cm}
		\includegraphics[scale=0.56]{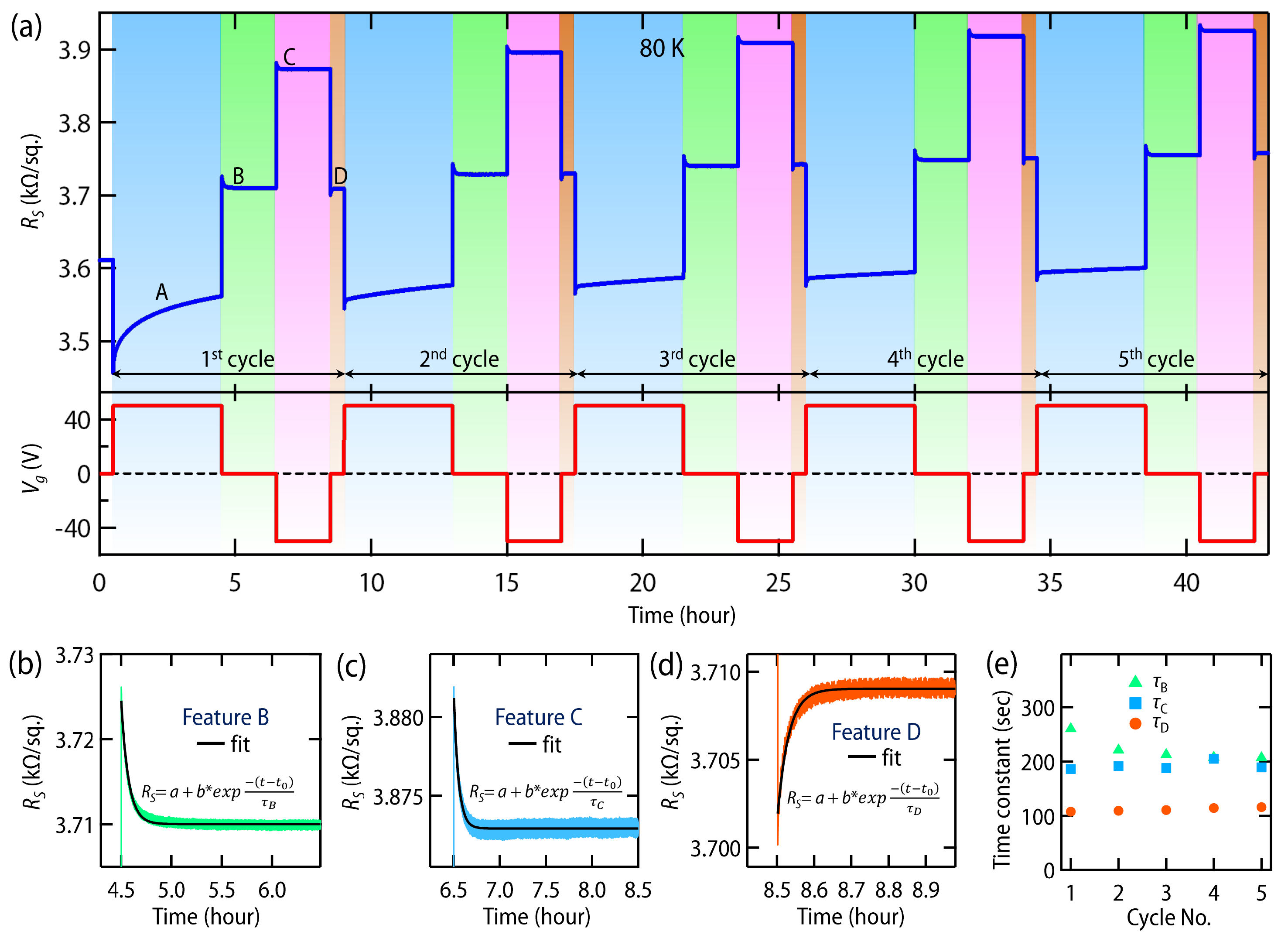}
		\caption{(a) Back gate voltage sweeping protocol (lower panel), and the corresponding variation of $R_S$ with time at 80 K (upper panel). (b,c,d) Magnified view of features B, C and D along with the fitting with single exponential function. Inset in the panels b-d shows the respective functional form of the formula used for fitting. $a$ and $b$ are the constants.  $t_0$ is the starting time of the features B, C and D. $\tau$$_\text{B}$, $\tau$$_\text{C}$ and $\tau$$_\text{D}$ denotes time constants associated with the features B, C and D respectively.  (e) Comparison of time constant of exponential features B, C $\&$ D and its variation with cycle.  } \label{fig:2}}
\end{figure*}

{\bf Evidence of charge trapping:}  We first discuss  the results of  gate voltage sweep measurements at 80 K. Before  each measurement, the sample was   heated to room temperature to remove all the trapped carriers from the previous gating history. Thereafter, it was cooled down to the desired temperature, and we waited for about 6 hours for temperature stabilization before starting any measurement. Since  +ve $V_{g}$ dopes electrons to the 2DEG, the sheet resistance ($R_{s}$) decreases  when $V_{g}$ was swept from 0 V to 50 V.   Surprisingly,  $R_{s}$ did not follow the same path when $V_g$ is swept backward to 0 V [see Fig.~\ref{fig:1}(d)].  The offset in $R_s$ between the forward and backward sweep, also observed for LAO/STO~\cite{biscaras:2014p6788,yin:2020p017702}, points to the trapping of some conduction electrons from the 2-DEG. Interestingly,  no such offset in $R_{s}$ was observed for -ve $V_g$ sweep (see Supplemental Material (SM)~\cite{sup}).  Furthermore, the irreversibility for +ve $V_g$ sweep is strongly dependent on the total time duration ($t_M$) for the entire forward and backward sweep.
    For example, left panel of Fig.~\ref{fig:1}(d) shows the variation of $R_s$ under a set of forward and backward sweep of $V_g$ (0 V $\rightarrow$ 20 V $\rightarrow$  5 V $\rightarrow$  30 V $\rightarrow$ 15 V $\rightarrow$ 40 V $\rightarrow$ 25 V $\rightarrow$ 50 V $\rightarrow$ 0 V), performed over a duration of 10 mins. All irreversible features become more prominent with an increase in $t_M$=  90 mins [middle panel of Fig.~\ref{fig:1}(d)] and $t_M$=240 mins [right panel of Fig.~\ref{fig:1}(d)].  Moreover, the value of $R_s$ at the maximum value of the applied $V_g$ (50 V) decreases with the increase of $t_M$. On the other hand, the value of $R_s$ after $V_g$ is reduced back to zero, increases with the increase of $t_M$.  All these observations  imply that the charge trapping processes  in GAO/STO heterostructures not only depend on the applied electric field but are also strongly time dependent.

{\bf Time dependent measurements with +ve and -ve step gate voltage:}  In order to understand the intriguing time-dependence of the charge trapping phenomena in GAO/STO heterostructures, we have designed a  gating protocol,  which consists of multiple cycles of +ve and -ve step gate voltages [shown in lower panel of Fig. \ref{fig:2}(a)] rather than a continuous sweeping of $V_g$. The duration of each step has been chosen to completely capture the concomitant trapping/detrapping processes. Upper panel of Fig. \ref{fig:2}(a) shows the corresponding variation of $R_S$ with applied $V_g$ at 80 K. In the first cycle, application of 0 $\rightarrow$ 50 V step gate voltage first leads to a sudden drop in $R_S$ due to the electrostatic charging, followed by a gradual increase of $R_S$ (called as feature A from now on), which  continues for several hours. Such slow relaxation of $R_S$ signifies charge trapping under +ve $V_g$, and understanding its microscopic origin is the primary focus of the present article.  Setting $V_g$ back to zero leads to a sudden jump in $R_S$ (similar to the discharging of a capacitor) followed by a decreasing $R_S$ (named as feature B for the rest of the paper).  In contrast to feature A, detrapping feature B involves a much smaller change in $R_S$, and saturates after few hundred seconds. Further application of negative gate voltage (0 V $\rightarrow$ -50 V) leads to   another fast detrapping (named   feature C), similar to feature B. Finally, upon setting the $V_g$ to zero, $R_S$ drops to a value, which is nearly equal to the value of $R_S$ at the end of feature B, followed by the emergence of a fast-rising trapping feature, denoted as D.   This entire voltage sweeping cycle has been repeated consecutively.

 While feature A has been observed earlier for the LAO/STO system~\cite{yin:2020p017702,biscaras:2014p6788}, features B, C and D are completely new. Interestingly, features  B, C and D can be fitted with a single exponential function [see Fig. \ref{fig:2}(b)-\ref{fig:2}(d)]. Since the time constants $\tau$$_\text{B}$, $\tau$$_\text{C}$ $\&$ $\tau$$_\text{D}$ associated with the features B, C $\&$ D, respectively are similar (a few hundred seconds), the microscopic origin of these three features must be the same. Also, these time constants do not change upon multiple cycling (Fig. \ref{fig:2}(e)).  It is important to note that feature C has also been independently observed by applying 0 $\rightarrow$ -50 V [see SM~\cite{sup}]. This clearly signifies that charge trapping (detrapping) under +ve (-ve) $V_g$ are independent processes and are not influenced by each other. The microscopic process behind these features will be discussed later in the text.

 \begin{figure}
	\centering{
		\hspace*{0 cm}
		\includegraphics[scale=.43
		]{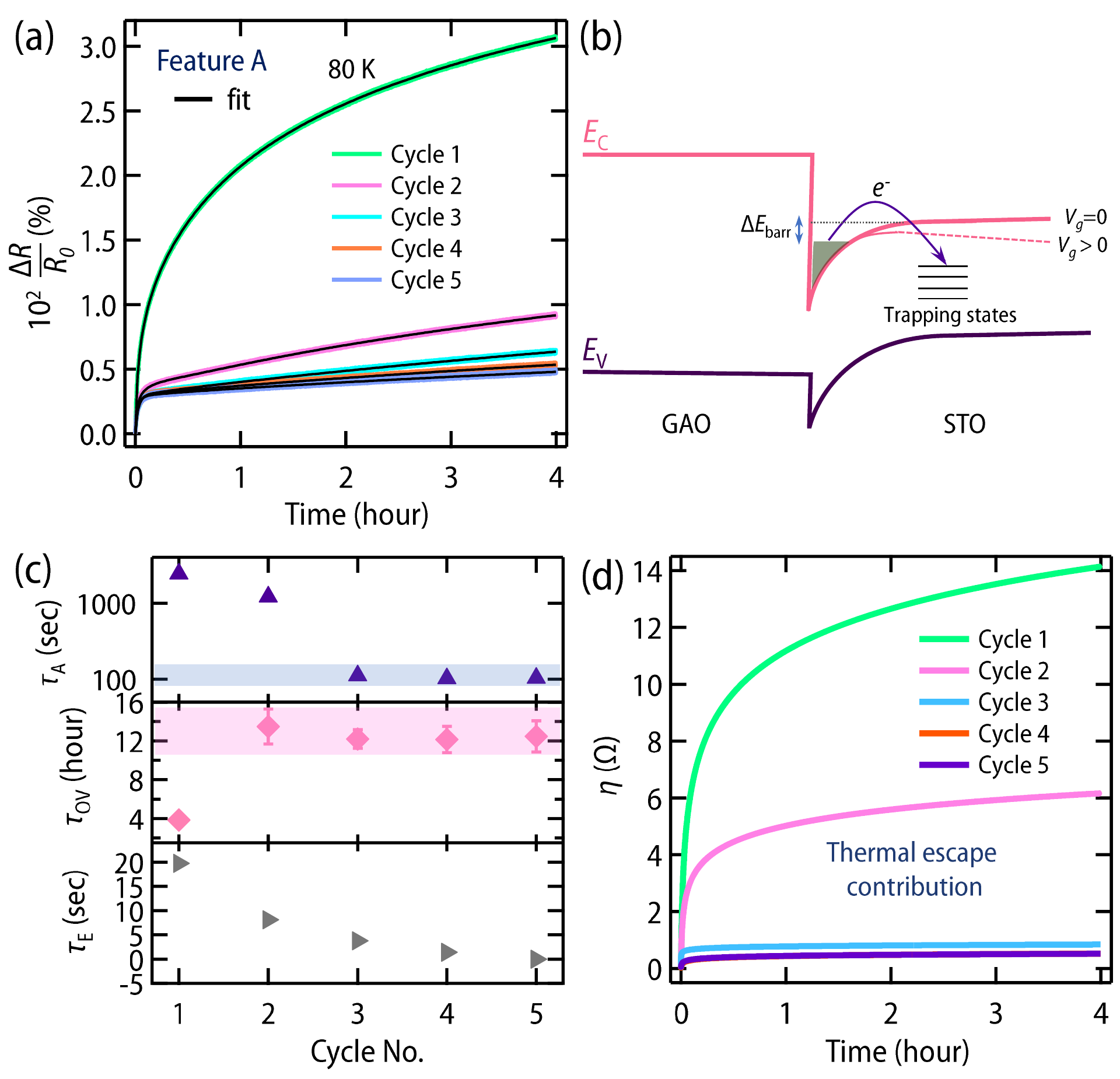}
		\caption{(a) Relative percentage change in resistance (10$^2$ $\frac{{\Delta}R}{R_0}$) of feature A due to charge trapping ($\Delta$$R$=$R$-$R_0$, where $R_0$ is the resistance just at the starting of the feature A). For comparison, starting time of the feature A for all the cycles have been manually shifted to zero. (b) A schematic to show thermal escape of electrons under +ve $V_g$. Application of +ve back gate voltage bends the conduction band and lowers the barrier height $\Delta$$E_\text{barr}$ for thermal escape. Application of -ve $V_g$ would bend the band upwards in energy which will increase $\Delta$$E_\text{barr}$ for thermal escape.  (c) Variation of $\tau$$_\text{A}$, $\tau_\text{OV}$ and $\tau$$_\text{E}$ upon multiple cycling.  (d) Variation of thermal escape contribution with increasing cycle number. } \label{fig:3}}
\end{figure}

In contrast to features B, C and D, feature A shows strong cycle dependence [Fig. \ref{fig:2}(a)]. This is more prominent in Fig. \ref{fig:3}(a), where we show relative percentage change in resistance (10$^2$ $\frac{\Delta R}{R_0}$) at 80 K for different cycles. There is a drastic change  from 1$^\text{st}$ to 2$^\text{nd}$ cycle and doesn't change much after that.  This observation  clearly emphasizes the presence  of multiple simultaneous processes responsible for charge trapping under +ve $V_g$ and further points to vanishing of one of the processes after the first cycle.  We could not describe the time dependence by considering either carrier trapping due to the clustering of OVs~\cite{yin:2020p017702} or thermal escape of carriers from quantum well~\cite{biscaras:2014p6788} or combination of both effect [see SM~\cite{sup}].
To capture feature A entirely, we consider the following function, which comprises of three simultaneous additive processes of charge trapping (also see SM~\cite{sup}).
\begin{equation}
R (t) = \alpha + \beta*e^{-\frac{t}{\tau_\text{A}}}+ \zeta*e^{-\frac{t}{\tau_\text{OV}}}+ f*ln(1+\frac{t}{\tau_\text{E}})  \label{eq:1}
\end{equation}
This hybrid function provides an excellent fit to feature A for all the cycles [Fig.\ref{fig:3}(a)], at different temperatures (shown later) and for different values of $V_g$ (also see SM~\cite{sup}).
The first exponential term accounts for a purely exponential process analogous to features B, C and D. It will be demonstrated later that this contribution is related  to the trapping of free carriers at the ferroelastic twin walls of STO. The second exponential term accounts for the trapping of free carriers in the mid gap states formed due to the electric field induced OV clusters~\cite{Seri:2013p125110,biscaras:2014p6788,yin:2020p017702}. The logarithmic term  represents loss of carriers due to the thermal escape of electrons from the quantum well~\cite{biscaras:2014p6788}, mediated by a downward band bending under +ve $V_g$ [Fig. \ref{fig:3}(b)]. $\alpha$, $\beta$, $ \zeta$ and $f$ are constants for a given voltage and temperature. $\tau_{\text{A}}$ and $\tau_{\text{OV}}$ are time constants associated with two exponential processes, whereas $\tau_{\text{E}}$ is the characteristic  time scale associated with thermal escape of carriers from the 2-DEG.

\begin{figure*}
	\centering{
		{~}\hspace*{-0cm}
		\includegraphics[scale=0.49]{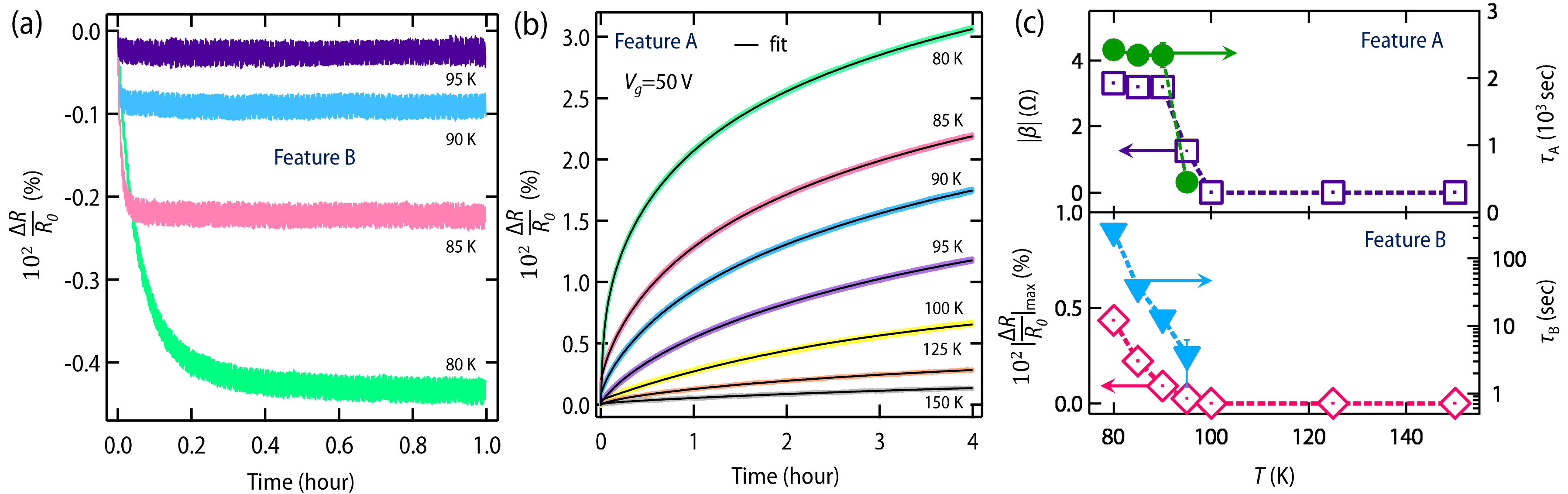}
		\caption{(a) Temperature dependence of feature B (first cycle).  (b) Temperature dependence of feature A (first cycle) for $V_g$ = 50 V. (c) Temperature evolution of $\tau_{A}$ and $|\beta|$ obtained from fitting of feature A (upper panel). Lower panel shows the temperature evolution of   $\tau_{B}$  obtained from fitting of feature B and  magnitude of maximum relative percentage change in resistance  (10$^2$ $\vert$$\frac{{\Delta}R}{R_0}$$\vert$$_\text{max}$) due to detrapping feature B. } \label{fig:4}}
\end{figure*}

Fig. \ref{fig:3}(c) shows the time constants (obtained from fitting) associated with the two exponential processes and thermal escape due to band bending (see  SM~\cite{sup} for cycle dependence of $\beta$ and $\zeta$). As clearly evident, time scale associated with the first exponential is in the order of a few hundred seconds ($\tau_{\text{A}}$) and other is in the order of a few hours ($\tau_{\text{OV}}$). Interestingly $\tau_{\text{A}}$ saturates at around 100 seconds after the  2$^\text{nd}$ cycle, which is very similar to the time scales associated with features B, C and D. This observation strongly points to the presence of a process analogous to features B, C and D contributing to charge trapping in feature A. In contrast to $\tau_{\text{A}}$, the time scale associated with second exponential ($\tau_{\text{OV}}$) lasts for several hours and can be attributed  to  the migration of OVs under electric field followed by their clustering~\cite{Christensen:2016p021602,Lei:2014p5554,Szot:2006p312,de:2012p174109,Hanzig:2013p024104,yin:2020p017702}. OV clustering leads to formation of mid-gap states, making it an active center for charge trapping.  The absence of  the OV cluster contribution for  -ve $V_g$ is related to the fact that  OVs move away from the interface i.e. towards the bulk of the STO  under -ve $V_g$. Since OV concentration is very little within the bulk of STO, the probability of clustering is very low under -ve $V_g$  (a schematic has been shown in  SM~\cite{sup}). In contrast to $\tau_{\text{A}}$ and $\tau_{\text{OV}}$, which saturates at higher cycles, $\tau_{\text{E}}$ decreases monotonically with increasing cycle and eventually vansihes at higher cycles.

 The  huge change in feature A from 1$^\text{st}$ to 3$^\text{rd}$ cycle is primarily related to  the  cycle dependence of thermal escape contribution ($\eta$=$f$$*ln$(1+$\frac{t}{\tau_\text{E}}$)), as shown in Fig. \ref{fig:3}(d) (also see  SM~\cite{sup}). $\eta$ becomes negligible after 2$^\text{nd}$ cycle. This is expected as the thermal escape is controlled by the energy barrier $\Delta$$E_\text{barr}$ (Fig. \ref{fig:3}b). Once the charges escape irreversibly from the quantum well in the first cycle, the Fermi level goes deeper into the well [Fig. \ref{fig:3}(b)]. This would increase $\Delta$$E_\text{barr}$, and therefore lower the probability of thermal escape in subsequent cycles.  The thermal escape contribution is also absent for the negative gate voltage  as the $\Delta$$E_\text{barr}$ increases due to the band bending  towards higher energy under -ve $V_g$ [Fig. \ref{fig:3}(b)].

\begin{figure*}
	\centering{
		\hspace*{0 cm}
		\includegraphics[scale=.38]{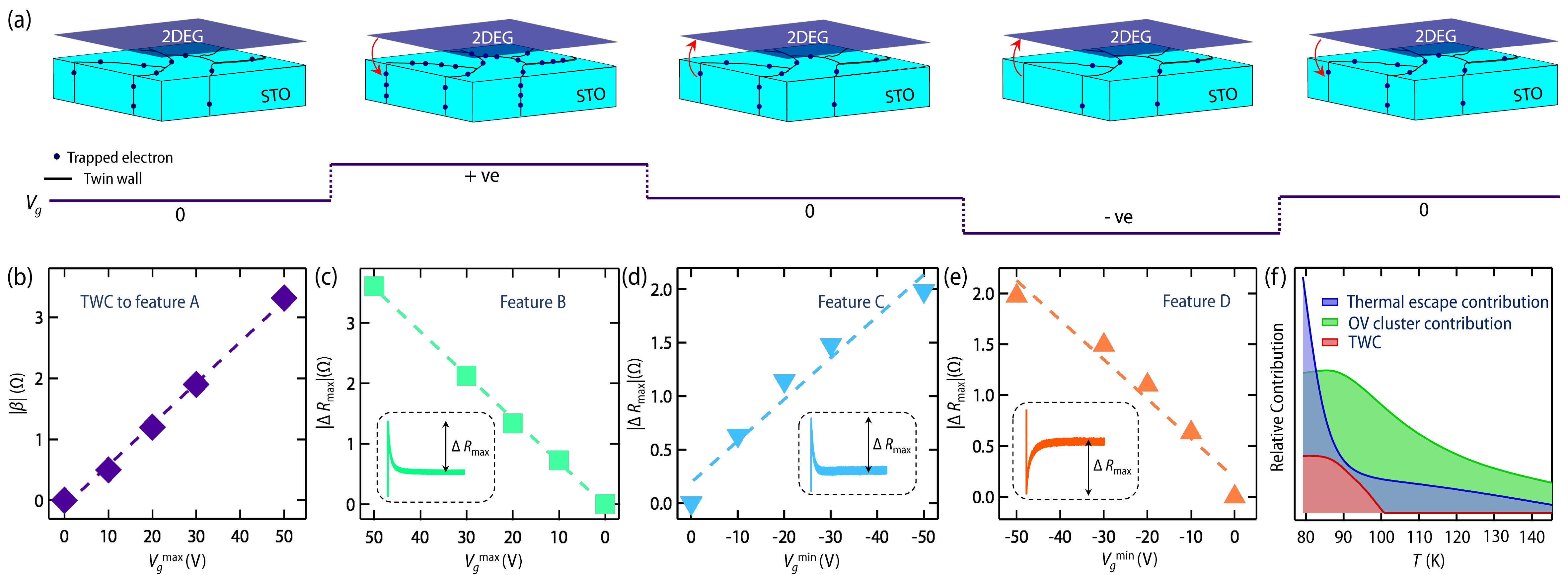}
		\caption{ (a) A set of schematics to show electric field induced trapping/detrapping of charges at twin walls of STO. 1$^\text{st}$ panel shows trapped electrons at twin walls of STO at zero back gate voltage. Application of 0 $\rightarrow$ +ve step $V_g$ leads to transfer of electrons from 2-DEG to twin walls (2$^\text{nd}$ panel). Switching off the field after 0 $\rightarrow$ +ve step $V_g$ leads to reversible transfer of electrons from twin walls to 2-DEG (3$^\text{rd}$ panel). Application of 0 $\rightarrow$ -ve step $V_g$ leads to transfer of electrons from twin walls to 2-DEG (4$^\text{th}$ panel). Switching off the voltage after application of 0 $\rightarrow$ -ve step $V_g$  leads to reversible transfer of electrons from 2-DEG to twin walls (5$^\text{th}$ panel). For visual clarity, STO substrate and 2-DEG have been shown physically separated from each other. (b) Variation of twin wall contribution to feature A in the 1$^\text{st}$ cycle at 80 K with ${V^\text{max}_{g}}$. (c) Variation of magnitude of maximum change in resistance due to detrapping feature B in the 1$^\text{st}$ cycle at 80 K with ${V^\text{max}_{g}}$. (d) Variation of magnitude of maximum change in resistance due to feature C in the 1$^\text{st}$ cycle at 80 K with ${V^\text{min}_{g}}$. (e) Variation of magnitude of maximum change in resistance due to feature D in the 1$^\text{st}$ cycle at 80 K with ${V^\text{min}_{g}}$. (f) A schematic depicting different charge trapping mechanisms present above and below the $T_c$ of STO. The change in resistance due to each of the three processes at the end of feature A (i.e. 4 hours after the application of a constant $V_g$ = 50 V) has been used to quantify their relative contributions. Dotted lines in the panels (b)-(e) denote fitting with a linear function.  Inset of panel (c)-(e) shows the definition of $\Delta$$R$$_\text{max}$ for the respective features.} \label{fig:5}}
\end{figure*}

{\bf Charge trapping in twin walls of SrTiO$_3$:}  In order to check how these  charge trapping/detrapping features evolve across the ferroelastic transition of STO, we have carried out similar measurements  at several temperatures.  We first discuss the variation of feature B, which has been captured by a single exponential function at 80 K. As evident from Fig. \ref{fig:4}(a) (first cycle of feature B),  this feature  gradually decreases with the increase of temperature, and  vanishes around 100 K,  which is very close to the ferroelastic phase transition temperature of STO. Similar temperature dependence has been also observed for feature C and D [see SM~\cite{sup}]. The feature A also gradually decreases with $T$ and becomes negligibly small above 150 K [Fig. \ref{fig:4}(b)]. The  analysis of this feature using  equation \ref{eq:1}, reveals that both, thermal escape ($\eta$) and OV cluster contribution term ($\zeta*e^{-\frac{t}{\tau_\text{OV}}}$) survives up to 150 K [see SM~\cite{sup}]. However, the contribution of the  $\beta*e^{-\frac{t}{\tau_\text{A}}}$ term of feature A vanishes around 100 K similar to features B, C and D. This is more prominent in Fig. \ref{fig:4}(c), where we have plotted $\tau_{\text{A}}$ and $|\beta|$ obtained from fitting of feature A   (upper panel) along with $\tau_{\text{B}}$ and 10$^2$ $\vert$$\frac{{\Delta}R}{R_0}$$\vert$$_\text{max}$ for feature B   (lower panel).   $|\beta|$ from feature A and 10$^2$ $\vert$$\frac{{\Delta}R}{R_0}$$\vert$$_\text{max}$ of feature B for the 2$^\text{nd}$ and  3$^\text{rd}$ cycles of measurements also vanishes at 100 K [see SM~\cite{sup}].

 Based on these experimental evidences, we propose the following microscopic processes below 100 K, responsible for the additional charge trapping contribution to feature A and the origin of features B, C and D. As mentioned earlier, polar twin walls, which appear spontaneously below the ferroelastic transition temperature  of STO, can trap mobile electrons even in the absence of any electric field [1$^\text{st}$ panel of Fig. \ref{fig:5}(a)]. Switching on the electric field can affect the amount of trapped charges at twin walls in two ways. It can either lead to the formation of new twin walls~\cite{Ma:2016p257601} and/or  can change the polarity of the twin walls. For the  first scenario, newly formed twin walls  would become additional source of trapping and can be the origin of the additional charge trapping contribution to feature A below 100 K. However, we rule out such a possibility in present case due to the following reasons.

 The existence of certain cutoff field for the formation of new twin walls was reported earlier by a scanning electron microscopy study~\cite{Ma:2016p257601}. More recent polarized optical microscopy study~\cite{Casals:2019p032025} further  demonstrated that even  electric field as high as 4 kV/cm is  not sufficient either to move twin  walls or to form new twin walls above 40 K (which is much lower than the range of temperatures investigated in present study). To check whether there is any cut off field for aditional charge trapping contribution to feature A in our system, we plot coefficient $|\beta|$ as a function of maximum value of step voltage (${V^\text{max}_{g}}$) in Fig. \ref{fig:5}(b) at 80 K (see  SM~\cite{sup}  for $V_g$ dependence of thermal escape and OV cluster contribution to charge trapping under +ve $V_g$). As evident, there is no cutoff voltage for the appearance of this additional contribution. This observation further strengthens our claim that the additional charge trapping contribution to feature A below $T_c$ of STO can not be attributed to electric field induced formation of new twin walls.

  Interestingly, the coefficient  $|\beta|$ scales almost linearly with maximum value of  $V_g$ (${V^\text{max}_{g}}$) for the applied +ve step voltage upto 50 V. At higher $V_g$, this dependence deviates from a linear behavior (see SM~\cite{sup}).  We also note that a previous scanning stress microscopy study demonstrated that the polarity of the twin wall changes linearly with applied stress at the twin wall~\cite{Frenkel:2017p1203}. We believe that due to coupling between the ferroelastic and ferroelectric order parameters in STO~\cite{Pesquera:2018p235701}, the same linear dependence is reflected in our electric field dependent measurement. This would mean that, upon application of +ve $V_g$, there is an increase in the polarity of twin walls. This would enhance the charge holding capacity of individual twin wall~\cite{Kalisky:2013p1091,Mokr:2007p094110} and leads to the additional charge trapping contribution ($\beta*e^{-\frac{t}{\tau_\text{A}}}$, will be called as twin wall contribution (TWC) from now on) to feature A below 100 K [2$^\text{nd}$ panel of Fig. \ref{fig:5}(a)].

  Removal of the +ve electric field would lead to the reversible transfer of electrons from charged twin walls to the 2-DEG [3$^\text{rd}$ panel of Fig. \ref{fig:5}(a)]. This would pump in few extra  carriers to the 2-DEG, leading to the decaying nature of feature B.  Similar to TWC to feature A, the magnitude of maximum change in resistance due to the detrapping feature B [$|\Delta R_\text{max}|$=$|R_\text{sat}$-$R_0|$, $R_\text{sat}$ is saturation value of resistance and $R_0$ is the resistance at manually shifted zero time [inset of Fig. \ref{fig:5}(c)] also shows the same linear dependence with the +ve voltage applied in previous step (${V^\text{max}_{g}}$) upto 50 V [Fig. \ref{fig:5}(c)]. Interestingly, the value of $|\beta|$ obtained from the fitting of feature A is very close to $|\Delta$$R$$_\text{max}|$ for feature B at each $V_g$.   This observation signifies that the charge trapping at twin walls of STO under the +ve $V_g$ is a completely reversible process [also see SM~\cite{sup}]. This is in sharp contrast to the completely irreversible nature of thermal escape and OV clustering contributions to charge trapping under +ve $V_g$.

  Application of -ve $V_g$ further transfers electrons from  twin walls to the 2-DEG [4$^\text{th}$ panel of Fig. \ref{fig:5}(a)], resulting in the decaying feature C.  Absence of any cut off field and almost linear dependence of  $|\Delta R_\text{max}|$ [see inset of Fig. \ref{fig:5}(d) for definition of $\Delta R_\text{max}$] on the minimum value of $V_g$ (${V^\text{min}_{g}}$) for the -ve step voltage [Fig. \ref{fig:5}(d)] indicates that the decreased polarity of twin wall under -ve $V_g$ is the microscopic origin   of feature C.    Switching off this negative voltage thereafter leads to a reversible transfer of electrons back to twin walls [5$^\text{th}$ panel of Fig. \ref{fig:5}(a)] and results in the appearance of the rising feature D. $|\Delta R_\text{max}|$ [see inset of Fig. \ref{fig:5}(e) for definition of $\Delta R_\text{max}$] for feature D also scales almost linearly with ${V^\text{min}_{g}}$ [Fig. \ref{fig:5}(e)].   We also note that the $|\Delta R_\text{max}|$ for features C and D are very similar, as expected for a reversible trapping/detrapping process. However, $|\Delta R_\text{max}|$ for feature C is lower than the corresponding $|\beta|$ for each $V_g$. Such a peculiar behavior can be attributed to   the multi-orbital nature of electronic bands near the Fermi level~\cite{zhang:2019p257601}.

In the present article, we have investigated different charge trapping/detrapping mechanisms across the structural phase transition temperature of STO. It is also important to note that the polar twin walls in STO undergo a series of phase transition from a domain glass phase to domain solid phase at lower temperatures~\cite{Kustov:2020p016801}. We expect similar glassy behavior to be reflected in the temporal relaxation of resistance (under applied electric field) in our samples at lower temperatures.

\section{Conclusions}

In conclusion, our detailed temperature and electric field dependent back gating experiments on GAO/STO heterostructure reveal that the charge trapping  in GAO/STO is quite complex and requires invoking simultaneous multiple trapping mechanisms [Fig. \ref{fig:5}(f)]. We detect clear signature of charge trapping (detrapping) at (from)  ferroelastic twin walls of STO substrate, below its structural transition temperature. Such charge trapping/detrapping at twin wall has not been reported  for any SrTiO$_3$ based heterostructures so far. This contribution for GAO/STO heterostructures is  completely reversible  with the cycling of positive and negative $V_g$. On the contrary, charge trapping contributions from the thermal escape of electrons from the quantum well, and the midgap state formation due to oxygen vacancy clustering are completely irreversible and present only under positive gate voltages.

 Our study highlights that these charge trapping/detrapping processes would  introduce time dependent features in  electrostatic gating experiments on  devices involving STO as a dielectric material and proper care must be taken in interpreting any data, especially in the case of studying memory effects. The present work also suggests that the scattering of conduction electrons at the STO's domain wall can be another limiting factor for the  electron mobility of STO-based 2-DEGs~\cite{Trier:2018p293002,Christensen:2018p054004}. Further, these twin walls can be treated as nanoscale local gate as the local carrier density can be changed by tuning the polarity. Since these twin walls can also be moved spatially by application of stress~\cite{Frenkel:2017p1203} and/or electric field~\cite{Casals:2019p032025}, they can be used to build memory elements~\cite{Salje:2020p164104,Nataf:2020p634} where polar twin walls would carry information similar to the racetrack memory technology~\cite{Parkin:190p190}.

\section{Acknowledgement}

 This work was funded by a DST Nanomission grant (DST/NM/NS/2018/246) and a  SERB Early Career Research Award (ECR/2018/001512). S.M. also acknowledges support from Infosys Foundation, Bangalore. The authors acknowledge AFM and XRD facilities at the Department of Physics, IISc Bangalore.


\begin{thebibliography}{0}%
\makeatletter
\providecommand \@ifxundefined [1]{%
 \@ifx{#1\undefined}
}%
\providecommand \@ifnum [1]{%
 \ifnum #1\expandafter \@firstoftwo
 \else \expandafter \@secondoftwo
 \fi
}%
\providecommand \@ifx [1]{%
 \ifx #1\expandafter \@firstoftwo
 \else \expandafter \@secondoftwo
 \fi
}%
\providecommand \natexlab [1]{#1}%
\providecommand \enquote  [1]{``#1''}%
\providecommand \bibnamefont  [1]{#1}%
\providecommand \bibfnamefont [1]{#1}%
\providecommand \citenamefont [1]{#1}%
\providecommand \href@noop [0]{\@secondoftwo}%
\providecommand \href [0]{\begingroup \@sanitize@url \@href}%
\providecommand \@href[1]{\@@startlink{#1}\@@href}%
\providecommand \@@href[1]{\endgroup#1\@@endlink}%
\providecommand \@sanitize@url [0]{\catcode `\\12\catcode `\$12\catcode
  `\&12\catcode `\#12\catcode `\^12\catcode `\_12\catcode `\%12\relax}%
\providecommand \@@startlink[1]{}%
\providecommand \@@endlink[0]{}%
\providecommand \url  [0]{\begingroup\@sanitize@url \@url }%
\providecommand \@url [1]{\endgroup\@href {#1}{\urlprefix }}%
\providecommand \urlprefix  [0]{URL }%
\providecommand \Eprint [0]{\href }%
\providecommand \doibase [0]{https://doi.org/}%
\providecommand \selectlanguage [0]{\@gobble}%
\providecommand \bibinfo  [0]{\@secondoftwo}%
\providecommand \bibfield  [0]{\@secondoftwo}%
\providecommand \translation [1]{[#1]}%
\providecommand \BibitemOpen [0]{}%
\providecommand \bibitemStop [0]{}%
\providecommand \bibitemNoStop [0]{.\EOS\space}%
\providecommand \EOS [0]{\spacefactor3000\relax}%
\providecommand \BibitemShut  [1]{\csname bibitem#1\endcsname}%
\let\auto@bib@innerbib\@empty
\end{thebibliography}%


\begin{thebibliography}{57}%
\makeatletter
\providecommand \@ifxundefined [1]{%
 \@ifx{#1\undefined}
}%
\providecommand \@ifnum [1]{%
 \ifnum #1\expandafter \@firstoftwo
 \else \expandafter \@secondoftwo
 \fi
}%
\providecommand \@ifx [1]{%
 \ifx #1\expandafter \@firstoftwo
 \else \expandafter \@secondoftwo
 \fi
}%
\providecommand \natexlab [1]{#1}%
\providecommand \enquote  [1]{``#1''}%
\providecommand \bibnamefont  [1]{#1}%
\providecommand \bibfnamefont [1]{#1}%
\providecommand \citenamefont [1]{#1}%
\providecommand \href@noop [0]{\@secondoftwo}%
\providecommand \href [0]{\begingroup \@sanitize@url \@href}%
\providecommand \@href[1]{\@@startlink{#1}\@@href}%
\providecommand \@@href[1]{\endgroup#1\@@endlink}%
\providecommand \@sanitize@url [0]{\catcode `\\12\catcode `\$12\catcode
  `\&12\catcode `\#12\catcode `\^12\catcode `\_12\catcode `\%12\relax}%
\providecommand \@@startlink[1]{}%
\providecommand \@@endlink[0]{}%
\providecommand \url  [0]{\begingroup\@sanitize@url \@url }%
\providecommand \@url [1]{\endgroup\@href {#1}{\urlprefix }}%
\providecommand \urlprefix  [0]{URL }%
\providecommand \Eprint [0]{\href }%
\providecommand \doibase [0]{https://doi.org/}%
\providecommand \selectlanguage [0]{\@gobble}%
\providecommand \bibinfo  [0]{\@secondoftwo}%
\providecommand \bibfield  [0]{\@secondoftwo}%
\providecommand \translation [1]{[#1]}%
\providecommand \BibitemOpen [0]{}%
\providecommand \bibitemStop [0]{}%
\providecommand \bibitemNoStop [0]{.\EOS\space}%
\providecommand \EOS [0]{\spacefactor3000\relax}%
\providecommand \BibitemShut  [1]{\csname bibitem#1\endcsname}%
\let\auto@bib@innerbib\@empty
\bibitem [{\citenamefont {Ahn}\ \emph {et~al.}(2003)\citenamefont {Ahn},
  \citenamefont {Triscone},\ and\ \citenamefont {Mannhart}}]{Ahn:2003p1015}%
  \BibitemOpen
  \bibfield  {author} {\bibinfo {author} {\bibfnamefont {C.}~\bibnamefont
  {Ahn}}, \bibinfo {author} {\bibfnamefont {J.-M.}\ \bibnamefont {Triscone}},\
  and\ \bibinfo {author} {\bibfnamefont {J.}~\bibnamefont {Mannhart}},\
  }\bibfield  {title} {\bibinfo {title} {Electric field effect in correlated
  oxide systems},\ }\href {https://doi.org/10.1038/nature01878} {\bibfield  {journal} {\bibinfo  {journal}
  {Nature}\ }\textbf {\bibinfo {volume} {424}},\ \bibinfo {pages} {1015}
  (\bibinfo {year} {2003})}\BibitemShut {NoStop}%
\bibitem [{\citenamefont {Vandenberghe}\ and\ \citenamefont
  {Fischetti}(2017)}]{Yandenberghe:2017p1}%
  \BibitemOpen
  \bibfield  {author} {\bibinfo {author} {\bibfnamefont {W.~G.}\ \bibnamefont
  {Vandenberghe}}\ and\ \bibinfo {author} {\bibfnamefont {M.~V.}\ \bibnamefont
  {Fischetti}},\ }\bibfield  {title} {\bibinfo {title} {Imperfect
  two-dimensional topological insulator field-effect transistors},\ }\href
  {https://doi.org/10.1038/ncomms14184} {\bibfield  {journal} {\bibinfo  {journal} {Nature communications}\
  }\textbf {\bibinfo {volume} {8}},\ \bibinfo {pages} {1} (\bibinfo {year}
  {2017})}\BibitemShut {NoStop}%
\bibitem [{\citenamefont {Collins}\ \emph {et~al.}(2018)\citenamefont
  {Collins}, \citenamefont {Tadich}, \citenamefont {Wu}, \citenamefont {Gomes},
  \citenamefont {Rodrigues}, \citenamefont {Liu}, \citenamefont {Hellerstedt},
  \citenamefont {Ryu}, \citenamefont {Tang}, \citenamefont {Mo} \emph
  {et~al.}}]{collins:2018p390}%
  \BibitemOpen
  \bibfield  {author} {\bibinfo {author} {\bibfnamefont {J.~L.}\ \bibnamefont
  {Collins}}, \bibinfo {author} {\bibfnamefont {A.}~\bibnamefont {Tadich}},
  \bibinfo {author} {\bibfnamefont {W.}~\bibnamefont {Wu}}, \bibinfo {author}
  {\bibfnamefont {L.~C.}\ \bibnamefont {Gomes}}, \bibinfo {author}
  {\bibfnamefont {J.~N.}\ \bibnamefont {Rodrigues}}, \bibinfo {author}
  {\bibfnamefont {C.}~\bibnamefont {Liu}}, \bibinfo {author} {\bibfnamefont
  {J.}~\bibnamefont {Hellerstedt}}, \bibinfo {author} {\bibfnamefont
  {H.}~\bibnamefont {Ryu}}, \bibinfo {author} {\bibfnamefont {S.}~\bibnamefont
  {Tang}}, \bibinfo {author} {\bibfnamefont {S.-K.}\ \bibnamefont {Mo}}, \emph
  {et~al.},\ }\bibfield  {title} {\bibinfo {title} {Electric-field-tuned
  topological phase transition in ultrathin Na$_3$Bi},\ }\href {https://doi.org/10.1038/s41586-018-0788-5}
  {\bibfield  {journal} {\bibinfo  {journal} {Nature}\ }\textbf {\bibinfo
  {volume} {564}},\ \bibinfo {pages} {390} (\bibinfo {year}
  {2018})}\BibitemShut {NoStop}%
\bibitem [{\citenamefont {Ohtomo}\ and\ \citenamefont
  {Hwang}(2004)}]{ohtomo:2004p423}%
  \BibitemOpen
  \bibfield  {author} {\bibinfo {author} {\bibfnamefont {A.}~\bibnamefont
  {Ohtomo}}\ and\ \bibinfo {author} {\bibfnamefont {H.}~\bibnamefont {Hwang}},\
  }\bibfield  {title} {\bibinfo {title} {A high-mobility electron gas at the
  LaAlO$_3$/SrTiO$_3$ heterointerface},\ }\href {https://doi.org/10.1038/nature02308} {\bibfield  {journal}
  {\bibinfo  {journal} {Nature}\ }\textbf {\bibinfo {volume} {427}},\ \bibinfo
  {pages} {423} (\bibinfo {year} {2004})}\BibitemShut {NoStop}%
\bibitem [{\citenamefont {Mannhart}\ and\ \citenamefont
  {Schlom}(2010)}]{Mannhart:2010p1607}%
  \BibitemOpen
  \bibfield  {author} {\bibinfo {author} {\bibfnamefont {J.}~\bibnamefont
  {Mannhart}}\ and\ \bibinfo {author} {\bibfnamefont {D.~G.}\ \bibnamefont
  {Schlom}},\ }\bibfield  {title} {\bibinfo {title} {Oxide interfaces--an
  opportunity for electronics},\ }\href
  {https://doi.org/10.1126/science.1181862} {\bibfield  {journal} {\bibinfo
  {journal} {Science}\ }\textbf {\bibinfo {volume} {327}},\ \bibinfo {pages}
  {1607} (\bibinfo {year} {2010})}\BibitemShut {NoStop}%
\bibitem [{\citenamefont {Mannhart}\ \emph {et~al.}(2008)\citenamefont
  {Mannhart}, \citenamefont {Blank}, \citenamefont {Hwang}, \citenamefont
  {Millis},\ and\ \citenamefont {Triscone}}]{mannhart:2008p1027}%
  \BibitemOpen
  \bibfield  {author} {\bibinfo {author} {\bibfnamefont {J.}~\bibnamefont
  {Mannhart}}, \bibinfo {author} {\bibfnamefont {D.~H.}\ \bibnamefont {Blank}},
  \bibinfo {author} {\bibfnamefont {H.}~\bibnamefont {Hwang}}, \bibinfo
  {author} {\bibfnamefont {A.}~\bibnamefont {Millis}},\ and\ \bibinfo {author}
  {\bibfnamefont {J.-M.}\ \bibnamefont {Triscone}},\ }\bibfield  {title}
  {\bibinfo {title} {Two-dimensional electron gases at oxide interfaces},\
  }\href {https://doi.org/10.1557/mrs2008.222} {\bibfield  {journal} {\bibinfo  {journal} {MRS bulletin}\
  }\textbf {\bibinfo {volume} {33}},\ \bibinfo {pages} {1027} (\bibinfo {year}
  {2008})}\BibitemShut {NoStop}%
\bibitem [{\citenamefont {Lee}\ \emph {et~al.}(2013)\citenamefont {Lee},
  \citenamefont {Xie}, \citenamefont {Sato}, \citenamefont {Bell},
  \citenamefont {Hikita}, \citenamefont {Hwang},\ and\ \citenamefont
  {Kao}}]{lee:2013p703}%
  \BibitemOpen
  \bibfield  {author} {\bibinfo {author} {\bibfnamefont {J.-S.}\ \bibnamefont
  {Lee}}, \bibinfo {author} {\bibfnamefont {Y.}~\bibnamefont {Xie}}, \bibinfo
  {author} {\bibfnamefont {H.}~\bibnamefont {Sato}}, \bibinfo {author}
  {\bibfnamefont {C.}~\bibnamefont {Bell}}, \bibinfo {author} {\bibfnamefont
  {Y.}~\bibnamefont {Hikita}}, \bibinfo {author} {\bibfnamefont
  {H.}~\bibnamefont {Hwang}},\ and\ \bibinfo {author} {\bibfnamefont {C.-C.}\
  \bibnamefont {Kao}},\ }\bibfield  {title} {\bibinfo {title} {Titanium \textit{d}$_{xy}$
  ferromagnetism at the LaAlO$_3$/SrTiO$_3$ interface},\ }\href {https://doi.org/10.1038/nmat3674} {\bibfield
  {journal} {\bibinfo  {journal} {Nature materials}\ }\textbf {\bibinfo
  {volume} {12}},\ \bibinfo {pages} {703} (\bibinfo {year} {2013})}\BibitemShut
  {NoStop}%
\bibitem [{\citenamefont {Brinkman}\ \emph {et~al.}(2007)\citenamefont
  {Brinkman}, \citenamefont {Huijben}, \citenamefont {Van~Zalk}, \citenamefont
  {Huijben}, \citenamefont {Zeitler}, \citenamefont {Maan}, \citenamefont
  {van~der Wiel}, \citenamefont {Rijnders}, \citenamefont {Blank},\ and\
  \citenamefont {Hilgenkamp}}]{brinkman:2007p493}%
  \BibitemOpen
  \bibfield  {author} {\bibinfo {author} {\bibfnamefont {A.}~\bibnamefont
  {Brinkman}}, \bibinfo {author} {\bibfnamefont {M.}~\bibnamefont {Huijben}},
  \bibinfo {author} {\bibfnamefont {M.}~\bibnamefont {Van~Zalk}}, \bibinfo
  {author} {\bibfnamefont {J.}~\bibnamefont {Huijben}}, \bibinfo {author}
  {\bibfnamefont {U.}~\bibnamefont {Zeitler}}, \bibinfo {author} {\bibfnamefont
  {J.}~\bibnamefont {Maan}}, \bibinfo {author} {\bibfnamefont {W.~G.}\
  \bibnamefont {van~der Wiel}}, \bibinfo {author} {\bibfnamefont
  {G.}~\bibnamefont {Rijnders}}, \bibinfo {author} {\bibfnamefont {D.~H.}\
  \bibnamefont {Blank}},\ and\ \bibinfo {author} {\bibfnamefont
  {H.}~\bibnamefont {Hilgenkamp}},\ }\bibfield  {title} {\bibinfo {title}
  {Magnetic effects at the interface between non-magnetic oxides},\ }\href
  {https://doi.org/10.1038/nmat1931} {\bibfield  {journal} {\bibinfo  {journal} {Nature materials}\ }\textbf
  {\bibinfo {volume} {6}},\ \bibinfo {pages} {493} (\bibinfo {year}
  {2007})}\BibitemShut {NoStop}%
\bibitem [{\citenamefont {Reyren}\ \emph {et~al.}(2007)\citenamefont {Reyren},
  \citenamefont {Thiel}, \citenamefont {Caviglia}, \citenamefont {Kourkoutis},
  \citenamefont {Hammerl}, \citenamefont {Richter}, \citenamefont {Schneider},
  \citenamefont {Kopp}, \citenamefont {R{\"u}etschi}, \citenamefont {Jaccard}
  \emph {et~al.}}]{reyren:2007p1196}%
  \BibitemOpen
  \bibfield  {author} {\bibinfo {author} {\bibfnamefont {N.}~\bibnamefont
  {Reyren}}, \bibinfo {author} {\bibfnamefont {S.}~\bibnamefont {Thiel}},
  \bibinfo {author} {\bibfnamefont {A.}~\bibnamefont {Caviglia}}, \bibinfo
  {author} {\bibfnamefont {L.~F.}\ \bibnamefont {Kourkoutis}}, \bibinfo
  {author} {\bibfnamefont {G.}~\bibnamefont {Hammerl}}, \bibinfo {author}
  {\bibfnamefont {C.}~\bibnamefont {Richter}}, \bibinfo {author} {\bibfnamefont
  {C.~W.}\ \bibnamefont {Schneider}}, \bibinfo {author} {\bibfnamefont
  {T.}~\bibnamefont {Kopp}}, \bibinfo {author} {\bibfnamefont {A.-S.}\
  \bibnamefont {R{\"u}etschi}}, \bibinfo {author} {\bibfnamefont
  {D.}~\bibnamefont {Jaccard}}, \emph {et~al.},\ }\bibfield  {title} {\bibinfo
  {title} {Superconducting interfaces between insulating oxides},\ }\href
  {https://doi.org/10.1126/science.1146006} {\bibfield  {journal} {\bibinfo  {journal} {Science}\ }\textbf {\bibinfo
  {volume} {317}},\ \bibinfo {pages} {1196} (\bibinfo {year}
  {2007})}\BibitemShut {NoStop}%
\bibitem [{\citenamefont {Stemmer}\ and\ \citenamefont
  {James~Allen}(2014)}]{stemmer:2014p151}%
  \BibitemOpen
  \bibfield  {author} {\bibinfo {author} {\bibfnamefont {S.}~\bibnamefont
  {Stemmer}}\ and\ \bibinfo {author} {\bibfnamefont {S.}~\bibnamefont
  {James~Allen}},\ }\bibfield  {title} {\bibinfo {title} {Two-dimensional
  electron gases at complex oxide interfaces},\ }\href {https://doi.org/10.1146/annurev-matsci-070813-113552} {\bibfield
  {journal} {\bibinfo  {journal} {Annual Review of Materials Research}\
  }\textbf {\bibinfo {volume} {44}},\ \bibinfo {pages} {151} (\bibinfo {year}
  {2014})}\BibitemShut {NoStop}%
\bibitem [{\citenamefont {King}\ \emph {et~al.}(2014)\citenamefont {King},
  \citenamefont {Walker}, \citenamefont {Tamai}, \citenamefont {De~La~Torre},
  \citenamefont {Eknapakul}, \citenamefont {Buaphet}, \citenamefont {Mo},
  \citenamefont {Meevasana}, \citenamefont {Bahramy},\ and\ \citenamefont
  {Baumberger}}]{king:2014p1}%
  \BibitemOpen
  \bibfield  {author} {\bibinfo {author} {\bibfnamefont {P.}~\bibnamefont
  {King}}, \bibinfo {author} {\bibfnamefont {S.~M.}\ \bibnamefont {Walker}},
  \bibinfo {author} {\bibfnamefont {A.}~\bibnamefont {Tamai}}, \bibinfo
  {author} {\bibfnamefont {A.}~\bibnamefont {De~La~Torre}}, \bibinfo {author}
  {\bibfnamefont {T.}~\bibnamefont {Eknapakul}}, \bibinfo {author}
  {\bibfnamefont {P.}~\bibnamefont {Buaphet}}, \bibinfo {author} {\bibfnamefont
  {S.-K.}\ \bibnamefont {Mo}}, \bibinfo {author} {\bibfnamefont
  {W.}~\bibnamefont {Meevasana}}, \bibinfo {author} {\bibfnamefont
  {M.}~\bibnamefont {Bahramy}},\ and\ \bibinfo {author} {\bibfnamefont
  {F.}~\bibnamefont {Baumberger}},\ }\bibfield  {title} {\bibinfo {title}
  {Quasiparticle dynamics and spin--orbital texture of the SrTiO$_3$
  two-dimensional electron gas},\ }\href {https://doi.org/10.1038/ncomms4414} {\bibfield  {journal}
  {\bibinfo  {journal} {Nature communications}\ }\textbf {\bibinfo {volume}
  {5}},\ \bibinfo {pages} {1} (\bibinfo {year} {2014})}\BibitemShut {NoStop}%
\bibitem [{\citenamefont {Vaz}\ \emph {et~al.}(2019)\citenamefont {Vaz},
  \citenamefont {No{\"e}l}, \citenamefont {Johansson}, \citenamefont
  {G{\"o}bel}, \citenamefont {Bruno}, \citenamefont {Singh}, \citenamefont
  {Mckeown-Walker}, \citenamefont {Trier}, \citenamefont {Vicente-Arche},
  \citenamefont {Sander} \emph {et~al.}}]{vaz:2019p1187}%
  \BibitemOpen
  \bibfield  {author} {\bibinfo {author} {\bibfnamefont {D.~C.}\ \bibnamefont
  {Vaz}}, \bibinfo {author} {\bibfnamefont {P.}~\bibnamefont {No{\"e}l}},
  \bibinfo {author} {\bibfnamefont {A.}~\bibnamefont {Johansson}}, \bibinfo
  {author} {\bibfnamefont {B.}~\bibnamefont {G{\"o}bel}}, \bibinfo {author}
  {\bibfnamefont {F.~Y.}\ \bibnamefont {Bruno}}, \bibinfo {author}
  {\bibfnamefont {G.}~\bibnamefont {Singh}}, \bibinfo {author} {\bibfnamefont
  {S.}~\bibnamefont {Mckeown-Walker}}, \bibinfo {author} {\bibfnamefont
  {F.}~\bibnamefont {Trier}}, \bibinfo {author} {\bibfnamefont {L.~M.}\
  \bibnamefont {Vicente-Arche}}, \bibinfo {author} {\bibfnamefont
  {A.}~\bibnamefont {Sander}}, \emph {et~al.},\ }\bibfield  {title} {\bibinfo
  {title} {Mapping spin--charge conversion to the band structure in a
  topological oxide two-dimensional electron gas},\ }\href {https://doi.org/10.1038/s41563-019-0467-4} {\bibfield
  {journal} {\bibinfo  {journal} {Nature materials}\ }\textbf {\bibinfo
  {volume} {18}},\ \bibinfo {pages} {1187} (\bibinfo {year}
  {2019})}\BibitemShut {NoStop}%
\bibitem [{\citenamefont {Ojha}\ \emph {et~al.}(2020)\citenamefont {Ojha},
  \citenamefont {Gogoi}, \citenamefont {Patidar}, \citenamefont {Patel},
  \citenamefont {Mandal}, \citenamefont {Kumar}, \citenamefont {Venkatesh},
  \citenamefont {Ganesan}, \citenamefont {Jain},\ and\ \citenamefont
  {Middey}}]{Ojha:2020p2000021}%
  \BibitemOpen
  \bibfield  {author} {\bibinfo {author} {\bibfnamefont {S.~K.}\ \bibnamefont
  {Ojha}}, \bibinfo {author} {\bibfnamefont {S.~K.}\ \bibnamefont {Gogoi}},
  \bibinfo {author} {\bibfnamefont {M.~M.}\ \bibnamefont {Patidar}}, \bibinfo
  {author} {\bibfnamefont {R.~K.}\ \bibnamefont {Patel}}, \bibinfo {author}
  {\bibfnamefont {P.}~\bibnamefont {Mandal}}, \bibinfo {author} {\bibfnamefont
  {S.}~\bibnamefont {Kumar}}, \bibinfo {author} {\bibfnamefont
  {R.}~\bibnamefont {Venkatesh}}, \bibinfo {author} {\bibfnamefont
  {V.}~\bibnamefont {Ganesan}}, \bibinfo {author} {\bibfnamefont
  {M.}~\bibnamefont {Jain}},\ and\ \bibinfo {author} {\bibfnamefont
  {S.}~\bibnamefont {Middey}},\ }\bibfield  {title} {\bibinfo {title} {Oxygen
  vacancy-induced topological hall effect in a nonmagnetic band insulator},\
  }\href {https://doi.org/10.1002/qute.202000021} {\bibfield  {journal}
  {\bibinfo  {journal} {Advanced Quantum Technologies}\ }\textbf {\bibinfo
  {volume} {3}},\ \bibinfo {pages} {2000021} (\bibinfo {year}
  {2020})}\BibitemShut {NoStop}%
\bibitem [{\citenamefont {Thiel}\ \emph {et~al.}(2006)\citenamefont {Thiel},
  \citenamefont {Hammerl}, \citenamefont {Schmehl}, \citenamefont {Schneider},\
  and\ \citenamefont {Mannhart}}]{thiel:2006p1942}%
  \BibitemOpen
  \bibfield  {author} {\bibinfo {author} {\bibfnamefont {S.}~\bibnamefont
  {Thiel}}, \bibinfo {author} {\bibfnamefont {G.}~\bibnamefont {Hammerl}},
  \bibinfo {author} {\bibfnamefont {A.}~\bibnamefont {Schmehl}}, \bibinfo
  {author} {\bibfnamefont {C.~W.}\ \bibnamefont {Schneider}},\ and\ \bibinfo
  {author} {\bibfnamefont {J.}~\bibnamefont {Mannhart}},\ }\bibfield  {title}
  {\bibinfo {title} {Tunable quasi-two-dimensional electron gases in oxide
  heterostructures},\ }\href {https://doi.org/10.1126/science.1131091} {\bibfield  {journal} {\bibinfo  {journal}
  {Science}\ }\textbf {\bibinfo {volume} {313}},\ \bibinfo {pages} {1942}
  (\bibinfo {year} {2006})}\BibitemShut {NoStop}%
\bibitem [{\citenamefont {Caviglia}\ \emph {et~al.}(2008)\citenamefont
  {Caviglia}, \citenamefont {Gariglio}, \citenamefont {Reyren}, \citenamefont
  {Jaccard}, \citenamefont {Schneider}, \citenamefont {Gabay}, \citenamefont
  {Thiel}, \citenamefont {Hammerl}, \citenamefont {Mannhart},\ and\
  \citenamefont {Triscone}}]{caviglia:2008p624}%
  \BibitemOpen
  \bibfield  {author} {\bibinfo {author} {\bibfnamefont {A.}~\bibnamefont
  {Caviglia}}, \bibinfo {author} {\bibfnamefont {S.}~\bibnamefont {Gariglio}},
  \bibinfo {author} {\bibfnamefont {N.}~\bibnamefont {Reyren}}, \bibinfo
  {author} {\bibfnamefont {D.}~\bibnamefont {Jaccard}}, \bibinfo {author}
  {\bibfnamefont {T.}~\bibnamefont {Schneider}}, \bibinfo {author}
  {\bibfnamefont {M.}~\bibnamefont {Gabay}}, \bibinfo {author} {\bibfnamefont
  {S.}~\bibnamefont {Thiel}}, \bibinfo {author} {\bibfnamefont
  {G.}~\bibnamefont {Hammerl}}, \bibinfo {author} {\bibfnamefont
  {J.}~\bibnamefont {Mannhart}},\ and\ \bibinfo {author} {\bibfnamefont
  {J.-M.}\ \bibnamefont {Triscone}},\ }\bibfield  {title} {\bibinfo {title}
  {Electric field control of the LaAlO$_3$/SrTiO$_3$ interface ground state},\
  }\href {https://doi.org/10.1038/nature07576} {\bibfield  {journal} {\bibinfo  {journal} {Nature}\ }\textbf
  {\bibinfo {volume} {456}},\ \bibinfo {pages} {624} (\bibinfo {year}
  {2008})}\BibitemShut {NoStop}%
\bibitem [{\citenamefont {Caviglia}\ \emph {et~al.}(2010)\citenamefont
  {Caviglia}, \citenamefont {Gabay}, \citenamefont {Gariglio}, \citenamefont
  {Reyren}, \citenamefont {Cancellieri},\ and\ \citenamefont
  {Triscone}}]{caviglia:2010p126803}%
  \BibitemOpen
  \bibfield  {author} {\bibinfo {author} {\bibfnamefont {A.}~\bibnamefont
  {Caviglia}}, \bibinfo {author} {\bibfnamefont {M.}~\bibnamefont {Gabay}},
  \bibinfo {author} {\bibfnamefont {S.}~\bibnamefont {Gariglio}}, \bibinfo
  {author} {\bibfnamefont {N.}~\bibnamefont {Reyren}}, \bibinfo {author}
  {\bibfnamefont {C.}~\bibnamefont {Cancellieri}},\ and\ \bibinfo {author}
  {\bibfnamefont {J.-M.}\ \bibnamefont {Triscone}},\ }\bibfield  {title}
  {\bibinfo {title} {Tunable rashba spin-orbit interaction at oxide
  interfaces},\ }\href {https://doi.org/10.1103/PhysRevLett.104.126803} {\bibfield  {journal} {\bibinfo  {journal}
  {Physical review letters}\ }\textbf {\bibinfo {volume} {104}},\ \bibinfo
  {pages} {126803} (\bibinfo {year} {2010})}\BibitemShut {NoStop}%
\bibitem [{\citenamefont {Bi}\ \emph {et~al.}(2014)\citenamefont {Bi},
  \citenamefont {Huang}, \citenamefont {Ryu}, \citenamefont {Lee},
  \citenamefont {Bark}, \citenamefont {Eom}, \citenamefont {Irvin},\ and\
  \citenamefont {Levy}}]{bi:2014p1}%
  \BibitemOpen
  \bibfield  {author} {\bibinfo {author} {\bibfnamefont {F.}~\bibnamefont
  {Bi}}, \bibinfo {author} {\bibfnamefont {M.}~\bibnamefont {Huang}}, \bibinfo
  {author} {\bibfnamefont {S.}~\bibnamefont {Ryu}}, \bibinfo {author}
  {\bibfnamefont {H.}~\bibnamefont {Lee}}, \bibinfo {author} {\bibfnamefont
  {C.-W.}\ \bibnamefont {Bark}}, \bibinfo {author} {\bibfnamefont {C.-B.}\
  \bibnamefont {Eom}}, \bibinfo {author} {\bibfnamefont {P.}~\bibnamefont
  {Irvin}},\ and\ \bibinfo {author} {\bibfnamefont {J.}~\bibnamefont {Levy}},\
  }\bibfield  {title} {\bibinfo {title} {Room-temperature
  electronically-controlled ferromagnetism at the LaAlO$_3$/SrTiO$_3$ interface},\
  }\href {https://doi.org/10.1038/ncomms6019} {\bibfield  {journal} {\bibinfo  {journal} {Nature
  communications}\ }\textbf {\bibinfo {volume} {5}},\ \bibinfo {pages} {1}
  (\bibinfo {year} {2014})}\BibitemShut {NoStop}%
\bibitem [{\citenamefont {Weaver}(1959)}]{Weaver:1959p274}%
  \BibitemOpen
  \bibfield  {author} {\bibinfo {author} {\bibfnamefont {H.}~\bibnamefont
  {Weaver}},\ }\bibfield  {title} {\bibinfo {title} {Dielectric properties of
  single crystals of SrTiO$_3$ at low temperatures},\ }\href
  {https://doi.org/https://doi.org/10.1016/0022-3697(59)90226-4} {\bibfield
  {journal} {\bibinfo  {journal} {Journal of Physics and Chemistry of Solids}\
  }\textbf {\bibinfo {volume} {11}},\ \bibinfo {pages} {274} (\bibinfo {year}
  {1959})}\BibitemShut {NoStop}%
\bibitem [{\citenamefont {Neville}\ \emph {et~al.}(1972)\citenamefont
  {Neville}, \citenamefont {Hoeneisen},\ and\ \citenamefont
  {Mead}}]{Neville:1972p2124}%
  \BibitemOpen
  \bibfield  {author} {\bibinfo {author} {\bibfnamefont {R.~C.}\ \bibnamefont
  {Neville}}, \bibinfo {author} {\bibfnamefont {B.}~\bibnamefont {Hoeneisen}},\
  and\ \bibinfo {author} {\bibfnamefont {C.~A.}\ \bibnamefont {Mead}},\
  }\bibfield  {title} {\bibinfo {title} {Permittivity of strontium titanate},\
  }\href {https://doi.org/10.1063/1.1661463} {\bibfield  {journal} {\bibinfo
  {journal} {Journal of Applied Physics}\ }\textbf {\bibinfo {volume} {43}},\
  \bibinfo {pages} {2124} (\bibinfo {year} {1972})},\  \BibitemShut {NoStop}%
\bibitem [{\citenamefont {Viana}\ \emph {et~al.}(1994)\citenamefont {Viana},
  \citenamefont {Lunkenheimer}, \citenamefont {Hemberger}, \citenamefont
  {B\"ohmer},\ and\ \citenamefont {Loidl}}]{Viana:1994p601}%
  \BibitemOpen
  \bibfield  {author} {\bibinfo {author} {\bibfnamefont {R.}~\bibnamefont
  {Viana}}, \bibinfo {author} {\bibfnamefont {P.}~\bibnamefont {Lunkenheimer}},
  \bibinfo {author} {\bibfnamefont {J.}~\bibnamefont {Hemberger}}, \bibinfo
  {author} {\bibfnamefont {R.}~\bibnamefont {B\"ohmer}},\ and\ \bibinfo
  {author} {\bibfnamefont {A.}~\bibnamefont {Loidl}},\ }\bibfield  {title}
  {\bibinfo {title} {Dielectric spectroscopy in SrTiO$_3$},\
  }\href {https://doi.org/10.1103/PhysRevB.50.601} {\bibfield  {journal}
  {\bibinfo  {journal} {Phys. Rev. B}\ }\textbf {\bibinfo {volume} {50}},\
  \bibinfo {pages} {601} (\bibinfo {year} {1994})}\BibitemShut {NoStop}%
\bibitem [{\citenamefont {Christensen}\ \emph {et~al.}(2016)\citenamefont
  {Christensen}, \citenamefont {Trier}, \citenamefont {von Soosten},
  \citenamefont {Prawiroatmodjo}, \citenamefont {Jespersen}, \citenamefont
  {Chen},\ and\ \citenamefont {Pryds}}]{Christensen:2016p021602}%
  \BibitemOpen
  \bibfield  {author} {\bibinfo {author} {\bibfnamefont {D.~V.}\ \bibnamefont
  {Christensen}}, \bibinfo {author} {\bibfnamefont {F.}~\bibnamefont {Trier}},
  \bibinfo {author} {\bibfnamefont {M.}~\bibnamefont {von Soosten}}, \bibinfo
  {author} {\bibfnamefont {G.~E. D.~K.}\ \bibnamefont {Prawiroatmodjo}},
  \bibinfo {author} {\bibfnamefont {T.~S.}\ \bibnamefont {Jespersen}}, \bibinfo
  {author} {\bibfnamefont {Y.~Z.}\ \bibnamefont {Chen}},\ and\ \bibinfo
  {author} {\bibfnamefont {N.}~\bibnamefont {Pryds}},\ }\bibfield  {title}
  {\bibinfo {title} {Electric field control of the $\gamma$-Al$_2$O$_3$/SrTiO$_3$
  interface conductivity at room temperature},\ }\href
  {https://doi.org/10.1063/1.4955490} {\bibfield  {journal} {\bibinfo
  {journal} {Applied Physics Letters}\ }\textbf {\bibinfo {volume} {109}},\
  \bibinfo {pages} {021602} (\bibinfo {year} {2016})}\BibitemShut {NoStop}%
\bibitem [{\citenamefont {Biscaras}\ \emph {et~al.}(2014)\citenamefont
  {Biscaras}, \citenamefont {Hurand}, \citenamefont {Feuillet-Palma},
  \citenamefont {Rastogi}, \citenamefont {Budhani}, \citenamefont {Reyren},
  \citenamefont {Lesne}, \citenamefont {Lesueur},\ and\ \citenamefont
  {Bergeal}}]{biscaras:2014p6788}%
  \BibitemOpen
  \bibfield  {author} {\bibinfo {author} {\bibfnamefont {J.}~\bibnamefont
  {Biscaras}}, \bibinfo {author} {\bibfnamefont {S.}~\bibnamefont {Hurand}},
  \bibinfo {author} {\bibfnamefont {C.}~\bibnamefont {Feuillet-Palma}},
  \bibinfo {author} {\bibfnamefont {A.}~\bibnamefont {Rastogi}}, \bibinfo
  {author} {\bibfnamefont {R.}~\bibnamefont {Budhani}}, \bibinfo {author}
  {\bibfnamefont {N.}~\bibnamefont {Reyren}}, \bibinfo {author} {\bibfnamefont
  {E.}~\bibnamefont {Lesne}}, \bibinfo {author} {\bibfnamefont
  {J.}~\bibnamefont {Lesueur}},\ and\ \bibinfo {author} {\bibfnamefont
  {N.}~\bibnamefont {Bergeal}},\ }\bibfield  {title} {\bibinfo {title} {Limit
  of the electrostatic doping in two-dimensional electron gases of LaXO$_3$ (X=
  Al, Ti)/SrTiO$_3$},\ }\href {https://doi.org/10.1038/srep06788} {\bibfield  {journal} {\bibinfo  {journal}
  {Scientific reports}\ }\textbf {\bibinfo {volume} {4}},\ \bibinfo {pages}
  {6788} (\bibinfo {year} {2014})}\BibitemShut {NoStop}%
\bibitem [{\citenamefont {Liu}\ \emph {et~al.}(2015)\citenamefont {Liu},
  \citenamefont {Gariglio}, \citenamefont {F{\^e}te}, \citenamefont {Li},
  \citenamefont {Boselli}, \citenamefont {Stornaiuolo},\ and\ \citenamefont
  {Triscone}}]{liu:2015p062805}%
  \BibitemOpen
  \bibfield  {author} {\bibinfo {author} {\bibfnamefont {W.}~\bibnamefont
  {Liu}}, \bibinfo {author} {\bibfnamefont {S.}~\bibnamefont {Gariglio}},
  \bibinfo {author} {\bibfnamefont {A.}~\bibnamefont {F{\^e}te}}, \bibinfo
  {author} {\bibfnamefont {D.}~\bibnamefont {Li}}, \bibinfo {author}
  {\bibfnamefont {M.}~\bibnamefont {Boselli}}, \bibinfo {author} {\bibfnamefont
  {D.}~\bibnamefont {Stornaiuolo}},\ and\ \bibinfo {author} {\bibfnamefont
  {J.-M.}\ \bibnamefont {Triscone}},\ }\bibfield  {title} {\bibinfo {title}
  {Magneto-transport study of top-and back-gated LaAlO$_3$/SrTiO$_3$
  heterostructures},\ }\href {https://doi.org/10.1063/1.4921068} {\bibfield  {journal} {\bibinfo  {journal}
  {APL materials}\ }\textbf {\bibinfo {volume} {3}},\ \bibinfo {pages} {062805}
  (\bibinfo {year} {2015})}\BibitemShut {NoStop}%
\bibitem [{\citenamefont {Yin}\ \emph {et~al.}(2020)\citenamefont {Yin},
  \citenamefont {Smink}, \citenamefont {Leermakers}, \citenamefont {Tang},
  \citenamefont {Lebedev}, \citenamefont {Zeitler}, \citenamefont {van~der
  Wiel}, \citenamefont {Hilgenkamp},\ and\ \citenamefont
  {Aarts}}]{yin:2020p017702}%
  \BibitemOpen
  \bibfield  {author} {\bibinfo {author} {\bibfnamefont {C.}~\bibnamefont
  {Yin}}, \bibinfo {author} {\bibfnamefont {A.~E.}\ \bibnamefont {Smink}},
  \bibinfo {author} {\bibfnamefont {I.}~\bibnamefont {Leermakers}}, \bibinfo
  {author} {\bibfnamefont {L.~M.}\ \bibnamefont {Tang}}, \bibinfo {author}
  {\bibfnamefont {N.}~\bibnamefont {Lebedev}}, \bibinfo {author} {\bibfnamefont
  {U.}~\bibnamefont {Zeitler}}, \bibinfo {author} {\bibfnamefont {W.~G.}\
  \bibnamefont {van~der Wiel}}, \bibinfo {author} {\bibfnamefont
  {H.}~\bibnamefont {Hilgenkamp}},\ and\ \bibinfo {author} {\bibfnamefont
  {J.}~\bibnamefont {Aarts}},\ }\bibfield  {title} {\bibinfo {title} {Electron
  trapping mechanism in LaAlO$_3$/SrTiO$_3$ heterostructures},\ }\href {https://doi.org/10.1103/PhysRevLett.124.017702}
  {\bibfield  {journal} {\bibinfo  {journal} {Physical Review Letters}\
  }\textbf {\bibinfo {volume} {124}},\ \bibinfo {pages} {017702} (\bibinfo
  {year} {2020})}\BibitemShut {NoStop}%
\bibitem [{\citenamefont {Bal}\ \emph {et~al.}(2017)\citenamefont {Bal},
  \citenamefont {Huang}, \citenamefont {Han}, \citenamefont {Ariando},
  \citenamefont {Venkatesan},\ and\ \citenamefont
  {Chandrasekhar}}]{Bal:2017p081604}%
  \BibitemOpen
  \bibfield  {author} {\bibinfo {author} {\bibfnamefont {V.~V.}\ \bibnamefont
  {Bal}}, \bibinfo {author} {\bibfnamefont {Z.}~\bibnamefont {Huang}}, \bibinfo
  {author} {\bibfnamefont {K.}~\bibnamefont {Han}}, \bibinfo {author}
  {\bibnamefont {Ariando}}, \bibinfo {author} {\bibfnamefont {T.}~\bibnamefont
  {Venkatesan}},\ and\ \bibinfo {author} {\bibfnamefont {V.}~\bibnamefont
  {Chandrasekhar}},\ }\bibfield  {title} {\bibinfo {title} {Electrostatic
  tuning of magnetism at the conducting (111) (La$_{0.3}$Sr$_{0.7}$)(Al$_{0.65}$Ta$_{0.35}$)/SrTiO$_3$
  interface},\ }\href {https://doi.org/10.1063/1.4986912} {\bibfield  {journal}
  {\bibinfo  {journal} {Applied Physics Letters}\ }\textbf {\bibinfo {volume}
  {111}},\ \bibinfo {pages} {081604} (\bibinfo {year} {2017})}\  \BibitemShut {NoStop}%
\bibitem [{\citenamefont {Chang}\ \emph {et~al.}(2014)\citenamefont {Chang},
  \citenamefont {Chu}, \citenamefont {Jeng}, \citenamefont {Cheng},
  \citenamefont {Lin}, \citenamefont {Yang},\ and\ \citenamefont
  {Chen}}]{Chang:2014p3522}%
  \BibitemOpen
  \bibfield  {author} {\bibinfo {author} {\bibfnamefont {C.-P.}\ \bibnamefont
  {Chang}}, \bibinfo {author} {\bibfnamefont {M.-W.}\ \bibnamefont {Chu}},
  \bibinfo {author} {\bibfnamefont {H.~T.}\ \bibnamefont {Jeng}}, \bibinfo
  {author} {\bibfnamefont {S.-L.}\ \bibnamefont {Cheng}}, \bibinfo {author}
  {\bibfnamefont {J.~G.}\ \bibnamefont {Lin}}, \bibinfo {author} {\bibfnamefont
  {J.-R.}\ \bibnamefont {Yang}},\ and\ \bibinfo {author} {\bibfnamefont
  {C.~H.}\ \bibnamefont {Chen}},\ }\bibfield  {title} {\bibinfo {title}
  {Condensation of two-dimensional oxide-interfacial charges into
  one-dimensional electron chains by the misfit-dislocation strain field},\
  }\href {https://doi.org/10.1038/ncomms4522} {\bibfield  {journal} {\bibinfo
  {journal} {Nature Communications}\ }\textbf {\bibinfo {volume} {5}},\
  \bibinfo {pages} {3522} (\bibinfo {year} {2014})}\BibitemShut {NoStop}%
\bibitem [{\citenamefont {Gunkel}\ \emph {et~al.}(2020)\citenamefont {Gunkel},
  \citenamefont {Christensen}, \citenamefont {Chen},\ and\ \citenamefont
  {Pryds}}]{Gunkel:2020p120505}%
  \BibitemOpen
  \bibfield  {author} {\bibinfo {author} {\bibfnamefont {F.}~\bibnamefont
  {Gunkel}}, \bibinfo {author} {\bibfnamefont {D.~V.}\ \bibnamefont
  {Christensen}}, \bibinfo {author} {\bibfnamefont {Y.~Z.}\ \bibnamefont
  {Chen}},\ and\ \bibinfo {author} {\bibfnamefont {N.}~\bibnamefont {Pryds}},\
  }\bibfield  {title} {\bibinfo {title} {Oxygen vacancies: The (in)visible
  friend of oxide electronics},\ }\href {https://doi.org/10.1063/1.5143309}
  {\bibfield  {journal} {\bibinfo  {journal} {Applied Physics Letters}\
  }\textbf {\bibinfo {volume} {116}},\ \bibinfo {pages} {120505} (\bibinfo
  {year} {2020})}\  \BibitemShut {NoStop}%
\bibitem [{\citenamefont {Cowley}(1964)}]{Cowley:1964pA981}%
  \BibitemOpen
  \bibfield  {author} {\bibinfo {author} {\bibfnamefont {R.~A.}\ \bibnamefont
  {Cowley}},\ }\bibfield  {title} {\bibinfo {title} {Lattice dynamics and phase
  transitions of strontium titanate},\ }\href
  {https://doi.org/10.1103/PhysRev.134.A981} {\bibfield  {journal} {\bibinfo
  {journal} {Phys. Rev.}\ }\textbf {\bibinfo {volume} {134}},\ \bibinfo {pages}
  {A981} (\bibinfo {year} {1964})}\BibitemShut {NoStop}%
\bibitem [{\citenamefont {Schiaffino}\ and\ \citenamefont
  {Stengel}(2017)}]{Schiaffino:2017p137601}%
  \BibitemOpen
  \bibfield  {author} {\bibinfo {author} {\bibfnamefont {A.}~\bibnamefont
  {Schiaffino}}\ and\ \bibinfo {author} {\bibfnamefont {M.}~\bibnamefont
  {Stengel}},\ }\bibfield  {title} {\bibinfo {title} {Macroscopic polarization
  from antiferrodistortive cycloids in ferroelastic SrTiO$_3$},\
  }\href {https://doi.org/10.1103/PhysRevLett.119.137601} {\bibfield  {journal}
  {\bibinfo  {journal} {Phys. Rev. Lett.}\ }\textbf {\bibinfo {volume} {119}},\
  \bibinfo {pages} {137601} (\bibinfo {year} {2017})}\BibitemShut {NoStop}%
\bibitem [{\citenamefont {Frenkel}\ \emph {et~al.}(2017)\citenamefont
  {Frenkel}, \citenamefont {Haham}, \citenamefont {Shperber}, \citenamefont
  {Bell}, \citenamefont {Xie}, \citenamefont {Chen}, \citenamefont {Hikita},
  \citenamefont {Hwang}, \citenamefont {Salje},\ and\ \citenamefont
  {Kalisky}}]{Frenkel:2017p1203}%
  \BibitemOpen
  \bibfield  {author} {\bibinfo {author} {\bibfnamefont {Y.}~\bibnamefont
  {Frenkel}}, \bibinfo {author} {\bibfnamefont {N.}~\bibnamefont {Haham}},
  \bibinfo {author} {\bibfnamefont {Y.}~\bibnamefont {Shperber}}, \bibinfo
  {author} {\bibfnamefont {C.}~\bibnamefont {Bell}}, \bibinfo {author}
  {\bibfnamefont {Y.}~\bibnamefont {Xie}}, \bibinfo {author} {\bibfnamefont
  {Z.}~\bibnamefont {Chen}}, \bibinfo {author} {\bibfnamefont {Y.}~\bibnamefont
  {Hikita}}, \bibinfo {author} {\bibfnamefont {H.~Y.}\ \bibnamefont {Hwang}},
  \bibinfo {author} {\bibfnamefont {E.~K.~H.}\ \bibnamefont {Salje}},\ and\
  \bibinfo {author} {\bibfnamefont {B.}~\bibnamefont {Kalisky}},\ }\bibfield
  {title} {\bibinfo {title} {Imaging and tuning polarity at srtio3
  domain walls},\ }\href {https://doi.org/10.1038/nmat4966} {\bibfield
  {journal} {\bibinfo  {journal} {Nature Materials}\ }\textbf {\bibinfo
  {volume} {16}},\ \bibinfo {pages} {1203} (\bibinfo {year}
  {2017})}\BibitemShut {NoStop}%
\bibitem [{\citenamefont {Salje}\ \emph {et~al.}(2013)\citenamefont {Salje},
  \citenamefont {Aktas}, \citenamefont {Carpenter}, \citenamefont {Laguta},\
  and\ \citenamefont {Scott}}]{Salje:2013p247603}%
  \BibitemOpen
  \bibfield  {author} {\bibinfo {author} {\bibfnamefont {E.~K.~H.}\
  \bibnamefont {Salje}}, \bibinfo {author} {\bibfnamefont {O.}~\bibnamefont
  {Aktas}}, \bibinfo {author} {\bibfnamefont {M.~A.}\ \bibnamefont
  {Carpenter}}, \bibinfo {author} {\bibfnamefont {V.~V.}\ \bibnamefont
  {Laguta}},\ and\ \bibinfo {author} {\bibfnamefont {J.~F.}\ \bibnamefont
  {Scott}},\ }\bibfield  {title} {\bibinfo {title} {Domains within domains and
  walls within walls: Evidence for polar domains in cryogenic
  SrTiO$_3$},\ }\href
  {https://doi.org/10.1103/PhysRevLett.111.247603} {\bibfield  {journal}
  {\bibinfo  {journal} {Phys. Rev. Lett.}\ }\textbf {\bibinfo {volume} {111}},\
  \bibinfo {pages} {247603} (\bibinfo {year} {2013})}\BibitemShut {NoStop}%
\bibitem [{\citenamefont {Zubko}\ \emph {et~al.}(2007)\citenamefont {Zubko},
  \citenamefont {Catalan}, \citenamefont {Buckley}, \citenamefont {Welche},\
  and\ \citenamefont {Scott}}]{Zubko:2007p167601}%
  \BibitemOpen
  \bibfield  {author} {\bibinfo {author} {\bibfnamefont {P.}~\bibnamefont
  {Zubko}}, \bibinfo {author} {\bibfnamefont {G.}~\bibnamefont {Catalan}},
  \bibinfo {author} {\bibfnamefont {A.}~\bibnamefont {Buckley}}, \bibinfo
  {author} {\bibfnamefont {P.~R.~L.}\ \bibnamefont {Welche}},\ and\ \bibinfo
  {author} {\bibfnamefont {J.~F.}\ \bibnamefont {Scott}},\ }\bibfield  {title}
  {\bibinfo {title} {Strain-gradient-induced polarization in
  SrTiO$_3$ single crystals},\ }\href
  {https://doi.org/10.1103/PhysRevLett.99.167601} {\bibfield  {journal}
  {\bibinfo  {journal} {Phys. Rev. Lett.}\ }\textbf {\bibinfo {volume} {99}},\
  \bibinfo {pages} {167601} (\bibinfo {year} {2007})}\BibitemShut {NoStop}%
\bibitem [{\citenamefont {Kalisky}\ \emph {et~al.}(2013)\citenamefont
  {Kalisky}, \citenamefont {Spanton}, \citenamefont {Noad}, \citenamefont
  {Kirtley}, \citenamefont {Nowack}, \citenamefont {Bell}, \citenamefont
  {Sato}, \citenamefont {Hosoda}, \citenamefont {Xie}, \citenamefont {Hikita},
  \citenamefont {Woltmann}, \citenamefont {Pfanzelt}, \citenamefont {Jany},
  \citenamefont {Richter}, \citenamefont {Hwang}, \citenamefont {Mannhart},\
  and\ \citenamefont {Moler}}]{Kalisky:2013p1091}%
  \BibitemOpen
  \bibfield  {author} {\bibinfo {author} {\bibfnamefont {B.}~\bibnamefont
  {Kalisky}}, \bibinfo {author} {\bibfnamefont {E.~M.}\ \bibnamefont
  {Spanton}}, \bibinfo {author} {\bibfnamefont {H.}~\bibnamefont {Noad}},
  \bibinfo {author} {\bibfnamefont {J.~R.}\ \bibnamefont {Kirtley}}, \bibinfo
  {author} {\bibfnamefont {K.~C.}\ \bibnamefont {Nowack}}, \bibinfo {author}
  {\bibfnamefont {C.}~\bibnamefont {Bell}}, \bibinfo {author} {\bibfnamefont
  {H.~K.}\ \bibnamefont {Sato}}, \bibinfo {author} {\bibfnamefont
  {M.}~\bibnamefont {Hosoda}}, \bibinfo {author} {\bibfnamefont
  {Y.}~\bibnamefont {Xie}}, \bibinfo {author} {\bibfnamefont {Y.}~\bibnamefont
  {Hikita}}, \bibinfo {author} {\bibfnamefont {C.}~\bibnamefont {Woltmann}},
  \bibinfo {author} {\bibfnamefont {G.}~\bibnamefont {Pfanzelt}}, \bibinfo
  {author} {\bibfnamefont {R.}~\bibnamefont {Jany}}, \bibinfo {author}
  {\bibfnamefont {C.}~\bibnamefont {Richter}}, \bibinfo {author} {\bibfnamefont
  {H.~Y.}\ \bibnamefont {Hwang}}, \bibinfo {author} {\bibfnamefont
  {J.}~\bibnamefont {Mannhart}},\ and\ \bibinfo {author} {\bibfnamefont
  {K.~A.}\ \bibnamefont {Moler}},\ }\bibfield  {title} {\bibinfo {title}
  {Locally enhanced conductivity due to the tetragonal domain structure in
  LaAlO$_3$/SrTiO$_3$ heterointerfaces},\ }\href {https://doi.org/10.1038/nmat3753}
  {\bibfield  {journal} {\bibinfo  {journal} {Nature Materials}\ }\textbf
  {\bibinfo {volume} {12}},\ \bibinfo {pages} {1091} (\bibinfo {year}
  {2013})}\BibitemShut {NoStop}%
\bibitem [{\citenamefont {Honig}\ \emph {et~al.}(2013)\citenamefont {Honig},
  \citenamefont {Sulpizio}, \citenamefont {Drori}, \citenamefont {Joshua},
  \citenamefont {Zeldov},\ and\ \citenamefont {Ilani}}]{Honig:2013p1112}%
  \BibitemOpen
  \bibfield  {author} {\bibinfo {author} {\bibfnamefont {M.}~\bibnamefont
  {Honig}}, \bibinfo {author} {\bibfnamefont {J.~A.}\ \bibnamefont {Sulpizio}},
  \bibinfo {author} {\bibfnamefont {J.}~\bibnamefont {Drori}}, \bibinfo
  {author} {\bibfnamefont {A.}~\bibnamefont {Joshua}}, \bibinfo {author}
  {\bibfnamefont {E.}~\bibnamefont {Zeldov}},\ and\ \bibinfo {author}
  {\bibfnamefont {S.}~\bibnamefont {Ilani}},\ }\bibfield  {title} {\bibinfo
  {title} {Local electrostatic imaging of striped domain order in
  LaAlO$_3$/SrTiO$_3$},\ }\href {https://doi.org/10.1038/nmat3810} {\bibfield
  {journal} {\bibinfo  {journal} {Nature Materials}\ }\textbf {\bibinfo
  {volume} {12}},\ \bibinfo {pages} {1112} (\bibinfo {year}
  {2013})}\BibitemShut {NoStop}%
\bibitem [{\citenamefont {Ma}\ \emph {et~al.}(2016)\citenamefont {Ma},
  \citenamefont {Scharinger}, \citenamefont {Zeng}, \citenamefont {Kohlberger},
  \citenamefont {Lange}, \citenamefont {St\"ohr}, \citenamefont {Wang},
  \citenamefont {Venkatesan}, \citenamefont {Kleiner}, \citenamefont {Scott},
  \citenamefont {Coey}, \citenamefont {Koelle},\ and\ \citenamefont
  {Ariando}}]{Ma:2016p257601}%
  \BibitemOpen
  \bibfield  {author} {\bibinfo {author} {\bibfnamefont {H.~J.~H.}\
  \bibnamefont {Ma}}, \bibinfo {author} {\bibfnamefont {S.}~\bibnamefont
  {Scharinger}}, \bibinfo {author} {\bibfnamefont {S.~W.}\ \bibnamefont
  {Zeng}}, \bibinfo {author} {\bibfnamefont {D.}~\bibnamefont {Kohlberger}},
  \bibinfo {author} {\bibfnamefont {M.}~\bibnamefont {Lange}}, \bibinfo
  {author} {\bibfnamefont {A.}~\bibnamefont {St\"ohr}}, \bibinfo {author}
  {\bibfnamefont {X.~R.}\ \bibnamefont {Wang}}, \bibinfo {author}
  {\bibfnamefont {T.}~\bibnamefont {Venkatesan}}, \bibinfo {author}
  {\bibfnamefont {R.}~\bibnamefont {Kleiner}}, \bibinfo {author} {\bibfnamefont
  {J.~F.}\ \bibnamefont {Scott}}, \bibinfo {author} {\bibfnamefont {J.~M.~D.}\
  \bibnamefont {Coey}}, \bibinfo {author} {\bibfnamefont {D.}~\bibnamefont
  {Koelle}},\ and\ \bibinfo {author} {\bibnamefont {Ariando}},\ }\bibfield
  {title} {\bibinfo {title} {Local electrical imaging of tetragonal domains and
  field-induced ferroelectric twin walls in conducting
  SrTiO$_3$},\ }\href
  {https://doi.org/10.1103/PhysRevLett.116.257601} {\bibfield  {journal}
  {\bibinfo  {journal} {Phys. Rev. Lett.}\ }\textbf {\bibinfo {volume} {116}},\
  \bibinfo {pages} {257601} (\bibinfo {year} {2016})}\BibitemShut {NoStop}%
\bibitem [{\citenamefont {Chen}\ \emph {et~al.}(2013)\citenamefont {Chen},
  \citenamefont {Bovet}, \citenamefont {Trier}, \citenamefont {Christensen},
  \citenamefont {Qu}, \citenamefont {Andersen}, \citenamefont {Kasama},
  \citenamefont {Zhang}, \citenamefont {Giraud}, \citenamefont {Dufouleur}
  \emph {et~al.}}]{chen:2013p16}%
  \BibitemOpen
  \bibfield  {author} {\bibinfo {author} {\bibfnamefont {Y.}~\bibnamefont
  {Chen}}, \bibinfo {author} {\bibfnamefont {N.}~\bibnamefont {Bovet}},
  \bibinfo {author} {\bibfnamefont {F.}~\bibnamefont {Trier}}, \bibinfo
  {author} {\bibfnamefont {D.}~\bibnamefont {Christensen}}, \bibinfo {author}
  {\bibfnamefont {F.}~\bibnamefont {Qu}}, \bibinfo {author} {\bibfnamefont
  {N.~H.}\ \bibnamefont {Andersen}}, \bibinfo {author} {\bibfnamefont
  {T.}~\bibnamefont {Kasama}}, \bibinfo {author} {\bibfnamefont
  {W.}~\bibnamefont {Zhang}}, \bibinfo {author} {\bibfnamefont
  {R.}~\bibnamefont {Giraud}}, \bibinfo {author} {\bibfnamefont
  {J.}~\bibnamefont {Dufouleur}}, \emph {et~al.},\ }\bibfield  {title}
  {\bibinfo {title} {A high-mobility two-dimensional electron gas at the
  spinel/perovskite interface of $\gamma$-Al$_2$O$_3$/SrTiO$_3$},\ }\href {https://doi.org/10.1038/ncomms2394}
  {\bibfield  {journal} {\bibinfo  {journal} {Nature communications}\ }\textbf
  {\bibinfo {volume} {4}},\ \bibinfo {pages} {1} (\bibinfo {year}
  {2013})}\BibitemShut {NoStop}%
\bibitem [{\citenamefont {Sch{\"u}tz}\ \emph {et~al.}(2017)\citenamefont
  {Sch{\"u}tz}, \citenamefont {Christensen}, \citenamefont {Borisov},
  \citenamefont {Pfaff}, \citenamefont {Scheiderer}, \citenamefont {Dudy},
  \citenamefont {Zapf}, \citenamefont {Gabel}, \citenamefont {Chen},
  \citenamefont {Pryds} \emph {et~al.}}]{schutz:2017p161409}%
  \BibitemOpen
  \bibfield  {author} {\bibinfo {author} {\bibfnamefont {P.}~\bibnamefont
  {Sch{\"u}tz}}, \bibinfo {author} {\bibfnamefont {D.~V.}\ \bibnamefont
  {Christensen}}, \bibinfo {author} {\bibfnamefont {V.}~\bibnamefont
  {Borisov}}, \bibinfo {author} {\bibfnamefont {F.}~\bibnamefont {Pfaff}},
  \bibinfo {author} {\bibfnamefont {P.}~\bibnamefont {Scheiderer}}, \bibinfo
  {author} {\bibfnamefont {L.}~\bibnamefont {Dudy}}, \bibinfo {author}
  {\bibfnamefont {M.}~\bibnamefont {Zapf}}, \bibinfo {author} {\bibfnamefont
  {J.}~\bibnamefont {Gabel}}, \bibinfo {author} {\bibfnamefont
  {Y.}~\bibnamefont {Chen}}, \bibinfo {author} {\bibfnamefont {N.}~\bibnamefont
  {Pryds}}, \emph {et~al.},\ }\bibfield  {title} {\bibinfo {title} {Microscopic
  origin of the mobility enhancement at a spinel/perovskite oxide
  heterointerface revealed by photoemission spectroscopy},\ }\href {https://doi.org/10.1103/PhysRevB.96.161409}
  {\bibfield  {journal} {\bibinfo  {journal} {Physical Review B}\ }\textbf
  {\bibinfo {volume} {96}},\ \bibinfo {pages} {161409} (\bibinfo {year}
  {2017})}\BibitemShut {NoStop}%
\bibitem [{\citenamefont {Christensen}\ \emph {et~al.}(2017)\citenamefont
  {Christensen}, \citenamefont {von Soosten}, \citenamefont {Trier},
  \citenamefont {Jespersen}, \citenamefont {Smith}, \citenamefont {Chen},\ and\
  \citenamefont {Pryds}}]{christensen:2017p1700026}%
  \BibitemOpen
  \bibfield  {author} {\bibinfo {author} {\bibfnamefont {D.~V.}\ \bibnamefont
  {Christensen}}, \bibinfo {author} {\bibfnamefont {M.}~\bibnamefont {von
  Soosten}}, \bibinfo {author} {\bibfnamefont {F.}~\bibnamefont {Trier}},
  \bibinfo {author} {\bibfnamefont {T.~S.}\ \bibnamefont {Jespersen}}, \bibinfo
  {author} {\bibfnamefont {A.}~\bibnamefont {Smith}}, \bibinfo {author}
  {\bibfnamefont {Y.}~\bibnamefont {Chen}},\ and\ \bibinfo {author}
  {\bibfnamefont {N.}~\bibnamefont {Pryds}},\ }\bibfield  {title} {\bibinfo
  {title} {Controlling the carrier density of SrTiO$_3$-based heterostructures
  with annealing},\ }\href {https://doi.org/10.1002/aelm.201700026} {\bibfield  {journal} {\bibinfo  {journal}
  {Advanced Electronic Materials}\ }\textbf {\bibinfo {volume} {3}},\ \bibinfo
  {pages} {1700026} (\bibinfo {year} {2017})}\BibitemShut {NoStop}%
\bibitem [{\citenamefont {Cao}\ \emph {et~al.}(2016)\citenamefont {Cao},
  \citenamefont {Liu}, \citenamefont {Shafer}, \citenamefont {Middey},
  \citenamefont {Meyers}, \citenamefont {Kareev}, \citenamefont {Zhong},
  \citenamefont {Kim}, \citenamefont {Ryan}, \citenamefont {Arenholz},\ and\
  \citenamefont {Chakhalian}}]{Cao:2016p1}%
  \BibitemOpen
  \bibfield  {author} {\bibinfo {author} {\bibfnamefont {Y.}~\bibnamefont
  {Cao}}, \bibinfo {author} {\bibfnamefont {X.}~\bibnamefont {Liu}}, \bibinfo
  {author} {\bibfnamefont {P.}~\bibnamefont {Shafer}}, \bibinfo {author}
  {\bibfnamefont {S.}~\bibnamefont {Middey}}, \bibinfo {author} {\bibfnamefont
  {D.}~\bibnamefont {Meyers}}, \bibinfo {author} {\bibfnamefont
  {M.}~\bibnamefont {Kareev}}, \bibinfo {author} {\bibfnamefont
  {Z.}~\bibnamefont {Zhong}}, \bibinfo {author} {\bibfnamefont {J.-W.}\
  \bibnamefont {Kim}}, \bibinfo {author} {\bibfnamefont {P.~J.}\ \bibnamefont
  {Ryan}}, \bibinfo {author} {\bibfnamefont {E.}~\bibnamefont {Arenholz}},\
  and\ \bibinfo {author} {\bibfnamefont {J.}~\bibnamefont {Chakhalian}},\
  }\bibfield  {title} {\bibinfo {title} {Anomalous orbital structure in a
  spinel{\textendash}perovskite interface},\ }\bibfield  {journal} {\bibinfo
  {journal} {npj Quantum Materials}\ }\textbf {\bibinfo {volume} {1}},\ \href
  {https://doi.org/10.1038/npjquantmats.2016.9} {10.1038/npjquantmats.2016.9}
  (\bibinfo {year} {2016})\BibitemShut {NoStop}%
\bibitem [{\citenamefont {Björck}\ and\ \citenamefont
  {Andersson}(2007)}]{Genxprogram}%
  \BibitemOpen
  \bibfield  {author} {\bibinfo {author} {\bibfnamefont {M.}~\bibnamefont
  {Björck}}\ and\ \bibinfo {author} {\bibfnamefont {G.}~\bibnamefont
  {Andersson}},\ }\bibfield  {title} {\bibinfo {title} {{GenX}: an extensible
  x-ray reflectivity refinement program utilizing differential evolution},\
  }\href {https://doi.org/10.1107/s0021889807045086} {\bibfield  {journal}
  {\bibinfo  {journal} {Journal of Applied Crystallography}\ }\textbf {\bibinfo
  {volume} {40}},\ \bibinfo {pages} {1174} (\bibinfo {year}
  {2007})}\BibitemShut {NoStop}%
\bibitem [{sup()}]{sup}%
  \BibitemOpen
  \href@noop {} {}\bibinfo {note} {See Supplemental Material for $R_S$ under -ve $V_g$ sweep, independent observation of feature C, functional form of feature A, cycle dependence of OV and twin wall
  contribution to feature A, schematic to explain electric field induced
  clustering of OVs and its redistribution during a complete voltage cycle,
  temperature dependence of feature A, temperature dependence of features C and
  D, temperature dependence of twin wall contribution to feature A and feature
  B in the 2$^\text{nd}$ and 3$^\text{rd}$ cycle, reversible nature of twin wall
  contribution to feature A under +ve $V_g$ and charge detrapping under -ve $V_g$,
  effect of electric field strength on charge trapping under +ve $V_g$, additional
  data on another 15 unit cell $\gamma$ -Al$_2$O$_3$/SrTiO$_3$ sample. It also
  includes reference number~\onlinecite{Sakudo:1971p851}.}\BibitemShut {Stop}%
\bibitem [{\citenamefont {Seri}\ \emph {et~al.}(2013)\citenamefont {Seri},
  \citenamefont {Schultz},\ and\ \citenamefont {Klein}}]{Seri:2013p125110}%
  \BibitemOpen
  \bibfield  {author} {\bibinfo {author} {\bibfnamefont {S.}~\bibnamefont
  {Seri}}, \bibinfo {author} {\bibfnamefont {M.}~\bibnamefont {Schultz}},\ and\
  \bibinfo {author} {\bibfnamefont {L.}~\bibnamefont {Klein}},\ }\bibfield
  {title} {\bibinfo {title} {Thermally activated recovery of electrical
  conductivity in LaAlO$_3$/SrTiO$_3$},\ }\href
  {https://doi.org/10.1103/PhysRevB.87.125110} {\bibfield  {journal} {\bibinfo
  {journal} {Phys. Rev. B}\ }\textbf {\bibinfo {volume} {87}},\ \bibinfo
  {pages} {125110} (\bibinfo {year} {2013})}\BibitemShut {NoStop}%
\bibitem [{\citenamefont {Lei}\ \emph {et~al.}(2014)\citenamefont {Lei},
  \citenamefont {Li}, \citenamefont {Chen}, \citenamefont {Xie}, \citenamefont
  {Chen}, \citenamefont {Wang}, \citenamefont {Wang}, \citenamefont {Shen},
  \citenamefont {Pryds}, \citenamefont {Hwang},\ and\ \citenamefont
  {Sun}}]{Lei:2014p5554}%
  \BibitemOpen
  \bibfield  {author} {\bibinfo {author} {\bibfnamefont {Y.}~\bibnamefont
  {Lei}}, \bibinfo {author} {\bibfnamefont {Y.}~\bibnamefont {Li}}, \bibinfo
  {author} {\bibfnamefont {Y.~Z.}\ \bibnamefont {Chen}}, \bibinfo {author}
  {\bibfnamefont {Y.~W.}\ \bibnamefont {Xie}}, \bibinfo {author} {\bibfnamefont
  {Y.~S.}\ \bibnamefont {Chen}}, \bibinfo {author} {\bibfnamefont {S.~H.}\
  \bibnamefont {Wang}}, \bibinfo {author} {\bibfnamefont {J.}~\bibnamefont
  {Wang}}, \bibinfo {author} {\bibfnamefont {B.~G.}\ \bibnamefont {Shen}},
  \bibinfo {author} {\bibfnamefont {N.}~\bibnamefont {Pryds}}, \bibinfo
  {author} {\bibfnamefont {H.~Y.}\ \bibnamefont {Hwang}},\ and\ \bibinfo
  {author} {\bibfnamefont {J.~R.}\ \bibnamefont {Sun}},\ }\bibfield  {title}
  {\bibinfo {title} {Visible-light-enhanced gating effect at the LaAlO$_3$/SrTiO$_3$
  interface},\ }\href {https://doi.org/10.1038/ncomms6554} {\bibfield
  {journal} {\bibinfo  {journal} {Nature Communications}\ }\textbf {\bibinfo
  {volume} {5}},\ \bibinfo {pages} {5554} (\bibinfo {year} {2014})}\BibitemShut
  {NoStop}%
\bibitem [{\citenamefont {Szot}\ \emph {et~al.}(2006)\citenamefont {Szot},
  \citenamefont {Speier}, \citenamefont {Bihlmayer},\ and\ \citenamefont
  {Waser}}]{Szot:2006p312}%
  \BibitemOpen
  \bibfield  {author} {\bibinfo {author} {\bibfnamefont {K.}~\bibnamefont
  {Szot}}, \bibinfo {author} {\bibfnamefont {W.}~\bibnamefont {Speier}},
  \bibinfo {author} {\bibfnamefont {G.}~\bibnamefont {Bihlmayer}},\ and\
  \bibinfo {author} {\bibfnamefont {R.}~\bibnamefont {Waser}},\ }\bibfield
  {title} {\bibinfo {title} {Switching the electrical resistance of individual
  dislocations in single-crystalline SrTiO$_3$},\ }\href
  {https://doi.org/10.1038/nmat1614} {\bibfield  {journal} {\bibinfo  {journal}
  {Nature Materials}\ }\textbf {\bibinfo {volume} {5}},\ \bibinfo {pages} {312}
  (\bibinfo {year} {2006})}\BibitemShut {NoStop}%
\bibitem [{\citenamefont {De~Souza}\ \emph {et~al.}(2012)\citenamefont
  {De~Souza}, \citenamefont {Metlenko}, \citenamefont {Park},\ and\
  \citenamefont {Weirich}}]{de:2012p174109}%
  \BibitemOpen
  \bibfield  {author} {\bibinfo {author} {\bibfnamefont {R.~A.}\ \bibnamefont
  {De~Souza}}, \bibinfo {author} {\bibfnamefont {V.}~\bibnamefont {Metlenko}},
  \bibinfo {author} {\bibfnamefont {D.}~\bibnamefont {Park}},\ and\ \bibinfo
  {author} {\bibfnamefont {T.~E.}\ \bibnamefont {Weirich}},\ }\bibfield
  {title} {\bibinfo {title} {Behavior of oxygen vacancies in single-crystal
  SrTiO$_3$: Equilibrium distribution and diffusion kinetics},\ }\href {https://doi.org/10.1103/PhysRevB.85.174109}
  {\bibfield  {journal} {\bibinfo  {journal} {Physical Review B}\ }\textbf
  {\bibinfo {volume} {85}},\ \bibinfo {pages} {174109} (\bibinfo {year}
  {2012})}\BibitemShut {NoStop}%
\bibitem [{\citenamefont {Hanzig}\ \emph {et~al.}(2013)\citenamefont {Hanzig},
  \citenamefont {Zschornak}, \citenamefont {Hanzig}, \citenamefont {Mehner},
  \citenamefont {St\"ocker}, \citenamefont {Abendroth}, \citenamefont
  {R\"oder}, \citenamefont {Talkenberger}, \citenamefont {Schreiber},
  \citenamefont {Rafaja}, \citenamefont {Gemming},\ and\ \citenamefont
  {Meyer}}]{Hanzig:2013p024104}%
  \BibitemOpen
  \bibfield  {author} {\bibinfo {author} {\bibfnamefont {J.}~\bibnamefont
  {Hanzig}}, \bibinfo {author} {\bibfnamefont {M.}~\bibnamefont {Zschornak}},
  \bibinfo {author} {\bibfnamefont {F.}~\bibnamefont {Hanzig}}, \bibinfo
  {author} {\bibfnamefont {E.}~\bibnamefont {Mehner}}, \bibinfo {author}
  {\bibfnamefont {H.}~\bibnamefont {St\"ocker}}, \bibinfo {author}
  {\bibfnamefont {B.}~\bibnamefont {Abendroth}}, \bibinfo {author}
  {\bibfnamefont {C.}~\bibnamefont {R\"oder}}, \bibinfo {author} {\bibfnamefont
  {A.}~\bibnamefont {Talkenberger}}, \bibinfo {author} {\bibfnamefont
  {G.}~\bibnamefont {Schreiber}}, \bibinfo {author} {\bibfnamefont
  {D.}~\bibnamefont {Rafaja}}, \bibinfo {author} {\bibfnamefont
  {S.}~\bibnamefont {Gemming}},\ and\ \bibinfo {author} {\bibfnamefont {D.~C.}\
  \bibnamefont {Meyer}},\ }\bibfield  {title} {\bibinfo {title}
  {Migration-induced field-stabilized polar phase in strontium titanate single
  crystals at room temperature},\ }\href
  {https://doi.org/10.1103/PhysRevB.88.024104} {\bibfield  {journal} {\bibinfo
  {journal} {Phys. Rev. B}\ }\textbf {\bibinfo {volume} {88}},\ \bibinfo
  {pages} {024104} (\bibinfo {year} {2013})}\BibitemShut {NoStop}%
\bibitem [{\citenamefont {Casals}\ \emph {et~al.}(2019)\citenamefont {Casals},
  \citenamefont {van Dijken}, \citenamefont {Herranz},\ and\ \citenamefont
  {Salje}}]{Casals:2019p032025}%
  \BibitemOpen
  \bibfield  {author} {\bibinfo {author} {\bibfnamefont {B.}~\bibnamefont
  {Casals}}, \bibinfo {author} {\bibfnamefont {S.}~\bibnamefont {van Dijken}},
  \bibinfo {author} {\bibfnamefont {G.}~\bibnamefont {Herranz}},\ and\ \bibinfo
  {author} {\bibfnamefont {E.~K.~H.}\ \bibnamefont {Salje}},\ }\bibfield
  {title} {\bibinfo {title} {Electric-field-induced avalanches and glassiness
  of mobile ferroelastic twin domains in cryogenic
  SrTiO$_3$},\ }\href
  {https://doi.org/10.1103/PhysRevResearch.1.032025} {\bibfield  {journal}
  {\bibinfo  {journal} {Phys. Rev. Research}\ }\textbf {\bibinfo {volume}
  {1}},\ \bibinfo {pages} {032025} (\bibinfo {year} {2019})}\BibitemShut
  {NoStop}%
\bibitem [{\citenamefont {Pesquera}\ \emph {et~al.}(2018)\citenamefont
  {Pesquera}, \citenamefont {Carpenter},\ and\ \citenamefont
  {Salje}}]{Pesquera:2018p235701}%
  \BibitemOpen
  \bibfield  {author} {\bibinfo {author} {\bibfnamefont {D.}~\bibnamefont
  {Pesquera}}, \bibinfo {author} {\bibfnamefont {M.~A.}\ \bibnamefont
  {Carpenter}},\ and\ \bibinfo {author} {\bibfnamefont {E.~K.~H.}\ \bibnamefont
  {Salje}},\ }\bibfield  {title} {\bibinfo {title} {Glasslike dynamics of polar
  domain walls in cryogenic SrTiO$_3$},\ }\href
  {https://doi.org/10.1103/PhysRevLett.121.235701} {\bibfield  {journal}
  {\bibinfo  {journal} {Phys. Rev. Lett.}\ }\textbf {\bibinfo {volume} {121}},\
  \bibinfo {pages} {235701} (\bibinfo {year} {2018})}\BibitemShut {NoStop}%
\bibitem [{\citenamefont {Mokr\'y}\ \emph {et~al.}(2007)\citenamefont
  {Mokr\'y}, \citenamefont {Tagantsev},\ and\ \citenamefont
  {Fousek}}]{Mokr:2007p094110}%
  \BibitemOpen
  \bibfield  {author} {\bibinfo {author} {\bibfnamefont {P.}~\bibnamefont
  {Mokr\'y}}, \bibinfo {author} {\bibfnamefont {A.~K.}\ \bibnamefont
  {Tagantsev}},\ and\ \bibinfo {author} {\bibfnamefont {J.}~\bibnamefont
  {Fousek}},\ }\bibfield  {title} {\bibinfo {title} {Pressure on charged domain
  walls and additional imprint mechanism in ferroelectrics},\ }\href
  {https://doi.org/10.1103/PhysRevB.75.094110} {\bibfield  {journal} {\bibinfo
  {journal} {Phys. Rev. B}\ }\textbf {\bibinfo {volume} {75}},\ \bibinfo
  {pages} {094110} (\bibinfo {year} {2007})}\BibitemShut {NoStop}%
\bibitem [{\citenamefont {Zhang}\ \emph {et~al.}(2019)\citenamefont {Zhang},
  \citenamefont {Lv}, \citenamefont {Zhang}, \citenamefont {Huang},
  \citenamefont {Wong}, \citenamefont {Yau}, \citenamefont {Chen},
  \citenamefont {Wen}, \citenamefont {Jiang}, \citenamefont {Zeng},
  \citenamefont {Hong},\ and\ \citenamefont {Dai}}]{zhang:2019p257601}%
  \BibitemOpen
  \bibfield  {author} {\bibinfo {author} {\bibfnamefont {F.}~\bibnamefont
  {Zhang}}, \bibinfo {author} {\bibfnamefont {P.}~\bibnamefont {Lv}}, \bibinfo
  {author} {\bibfnamefont {Y.}~\bibnamefont {Zhang}}, \bibinfo {author}
  {\bibfnamefont {S.}~\bibnamefont {Huang}}, \bibinfo {author} {\bibfnamefont
  {C.-M.}\ \bibnamefont {Wong}}, \bibinfo {author} {\bibfnamefont {H.-M.}\
  \bibnamefont {Yau}}, \bibinfo {author} {\bibfnamefont {X.}~\bibnamefont
  {Chen}}, \bibinfo {author} {\bibfnamefont {Z.}~\bibnamefont {Wen}}, \bibinfo
  {author} {\bibfnamefont {X.}~\bibnamefont {Jiang}}, \bibinfo {author}
  {\bibfnamefont {C.}~\bibnamefont {Zeng}}, \bibinfo {author} {\bibfnamefont
  {J.}~\bibnamefont {Hong}},\ and\ \bibinfo {author} {\bibfnamefont {J.-y.}\
  \bibnamefont {Dai}},\ }\bibfield  {title} {\bibinfo {title} {Modulating the
  electrical transport in the two-dimensional electron gas at
  LaAlO$_3$/SrTiO$_3$ heterostructures by interfacial
  flexoelectricity},\ }\href {https://doi.org/10.1103/PhysRevLett.122.257601}
  {\bibfield  {journal} {\bibinfo  {journal} {Phys. Rev. Lett.}\ }\textbf
  {\bibinfo {volume} {122}},\ \bibinfo {pages} {257601} (\bibinfo {year}
  {2019})}\BibitemShut {NoStop}%
\bibitem [{\citenamefont {Kustov}\ \emph {et~al.}(2020)\citenamefont {Kustov},
  \citenamefont {Liubimova},\ and\ \citenamefont {Salje}}]{Kustov:2020p016801}%
  \BibitemOpen
  \bibfield  {author} {\bibinfo {author} {\bibfnamefont {S.}~\bibnamefont
  {Kustov}}, \bibinfo {author} {\bibfnamefont {I.}~\bibnamefont {Liubimova}},\
  and\ \bibinfo {author} {\bibfnamefont {E.~K.~H.}\ \bibnamefont {Salje}},\
  }\bibfield  {title} {\bibinfo {title} {Domain dynamics in
  quantum-paraelectric SrTiO$_3$},\ }\href
  {https://doi.org/10.1103/PhysRevLett.124.016801} {\bibfield  {journal}
  {\bibinfo  {journal} {Phys. Rev. Lett.}\ }\textbf {\bibinfo {volume} {124}},\
  \bibinfo {pages} {016801} (\bibinfo {year} {2020})}\BibitemShut {NoStop}%
\bibitem [{\citenamefont {Trier}\ \emph {et~al.}(2018)\citenamefont {Trier},
  \citenamefont {Christensen},\ and\ \citenamefont
  {Pryds}}]{Trier:2018p293002}%
  \BibitemOpen
  \bibfield  {author} {\bibinfo {author} {\bibfnamefont {F.}~\bibnamefont
  {Trier}}, \bibinfo {author} {\bibfnamefont {D.~V.}\ \bibnamefont
  {Christensen}},\ and\ \bibinfo {author} {\bibfnamefont {N.}~\bibnamefont
  {Pryds}},\ }\bibfield  {title} {\bibinfo {title} {Electron mobility in oxide
  heterostructures},\ }\href {https://doi.org/10.1088/1361-6463/aac9aa}
  {\bibfield  {journal} {\bibinfo  {journal} {Journal of Physics D: Applied
  Physics}\ }\textbf {\bibinfo {volume} {51}},\ \bibinfo {pages} {293002}
  (\bibinfo {year} {2018})}\BibitemShut {NoStop}%
\bibitem [{\citenamefont {Christensen}\ \emph {et~al.}(2018)\citenamefont
  {Christensen}, \citenamefont {Frenkel}, \citenamefont {Sch\"utz},
  \citenamefont {Trier}, \citenamefont {Wissberg}, \citenamefont {Claessen},
  \citenamefont {Kalisky}, \citenamefont {Smith}, \citenamefont {Chen},\ and\
  \citenamefont {Pryds}}]{Christensen:2018p054004}%
  \BibitemOpen
  \bibfield  {author} {\bibinfo {author} {\bibfnamefont {D.~V.}\ \bibnamefont
  {Christensen}}, \bibinfo {author} {\bibfnamefont {Y.}~\bibnamefont
  {Frenkel}}, \bibinfo {author} {\bibfnamefont {P.}~\bibnamefont {Sch\"utz}},
  \bibinfo {author} {\bibfnamefont {F.}~\bibnamefont {Trier}}, \bibinfo
  {author} {\bibfnamefont {S.}~\bibnamefont {Wissberg}}, \bibinfo {author}
  {\bibfnamefont {R.}~\bibnamefont {Claessen}}, \bibinfo {author}
  {\bibfnamefont {B.}~\bibnamefont {Kalisky}}, \bibinfo {author} {\bibfnamefont
  {A.}~\bibnamefont {Smith}}, \bibinfo {author} {\bibfnamefont {Y.~Z.}\
  \bibnamefont {Chen}},\ and\ \bibinfo {author} {\bibfnamefont
  {N.}~\bibnamefont {Pryds}},\ }\bibfield  {title} {\bibinfo {title} {Electron
  mobility in $\gamma$-Al$_2$O$_3$/SrTiO$_3$},\
  }\href {https://doi.org/10.1103/PhysRevApplied.9.054004} {\bibfield
  {journal} {\bibinfo  {journal} {Phys. Rev. Applied}\ }\textbf {\bibinfo
  {volume} {9}},\ \bibinfo {pages} {054004} (\bibinfo {year}
  {2018})}\BibitemShut {NoStop}%
\bibitem [{\citenamefont {Salje}(2020)}]{Salje:2020p164104}%
  \BibitemOpen
  \bibfield  {author} {\bibinfo {author} {\bibfnamefont {E.~K.~H.}\
  \bibnamefont {Salje}},\ }\bibfield  {title} {\bibinfo {title} {Ferroelastic
  domain walls as templates for multiferroic devices},\ }\href
  {https://doi.org/10.1063/5.0029160} {\bibfield  {journal} {\bibinfo
  {journal} {Journal of Applied Physics}\ }\textbf {\bibinfo {volume} {128}},\
  \bibinfo {pages} {164104} (\bibinfo {year} {2020})},\  \BibitemShut {NoStop}%
\bibitem [{\citenamefont {Nataf}\ \emph {et~al.}(2020)\citenamefont {Nataf},
  \citenamefont {Guennou}, \citenamefont {Gregg}, \citenamefont {Meier},
  \citenamefont {Hlinka}, \citenamefont {Salje},\ and\ \citenamefont
  {Kreisel}}]{Nataf:2020p634}%
  \BibitemOpen
  \bibfield  {author} {\bibinfo {author} {\bibfnamefont {G.~F.}\ \bibnamefont
  {Nataf}}, \bibinfo {author} {\bibfnamefont {M.}~\bibnamefont {Guennou}},
  \bibinfo {author} {\bibfnamefont {J.~M.}\ \bibnamefont {Gregg}}, \bibinfo
  {author} {\bibfnamefont {D.}~\bibnamefont {Meier}}, \bibinfo {author}
  {\bibfnamefont {J.}~\bibnamefont {Hlinka}}, \bibinfo {author} {\bibfnamefont
  {E.~K.~H.}\ \bibnamefont {Salje}},\ and\ \bibinfo {author} {\bibfnamefont
  {J.}~\bibnamefont {Kreisel}},\ }\bibfield  {title} {\bibinfo {title}
  {Domain-wall engineering and topological defects in ferroelectric and
  ferroelastic materials},\ }\href {https://doi.org/10.1038/s42254-020-0235-z}
  {\bibfield  {journal} {\bibinfo  {journal} {Nature Reviews Physics}\ }\textbf
  {\bibinfo {volume} {2}},\ \bibinfo {pages} {634} (\bibinfo {year}
  {2020})}\BibitemShut {NoStop}%
\bibitem [{\citenamefont {Parkin}\ \emph {et~al.}(2008)\citenamefont {Parkin},
  \citenamefont {Hayashi},\ and\ \citenamefont {Thomas}}]{Parkin:190p190}%
  \BibitemOpen
  \bibfield  {author} {\bibinfo {author} {\bibfnamefont {S.~S.~P.}\
  \bibnamefont {Parkin}}, \bibinfo {author} {\bibfnamefont {M.}~\bibnamefont
  {Hayashi}},\ and\ \bibinfo {author} {\bibfnamefont {L.}~\bibnamefont
  {Thomas}},\ }\bibfield  {title} {\bibinfo {title} {Magnetic domain-wall
  racetrack memory},\ }\href {https://doi.org/10.1126/science.1145799}
  {\bibfield  {journal} {\bibinfo  {journal} {Science}\ }\textbf {\bibinfo
  {volume} {320}},\ \bibinfo {pages} {190} (\bibinfo {year} {2008})},\  \BibitemShut
  {NoStop}%
\bibitem [{\citenamefont {Sakudo}\ and\ \citenamefont
  {Unoki}(1971)}]{Sakudo:1971p851}%
  \BibitemOpen
  \bibfield  {author} {\bibinfo {author} {\bibfnamefont {T.}~\bibnamefont
  {Sakudo}}\ and\ \bibinfo {author} {\bibfnamefont {H.}~\bibnamefont {Unoki}},\
  }\bibfield  {title} {\bibinfo {title} {Dielectric properties of
  SrTiO$_3$ at low temperatures},\ }\href
  {https://doi.org/10.1103/PhysRevLett.26.851} {\bibfield  {journal} {\bibinfo
  {journal} {Phys. Rev. Lett.}\ }\textbf {\bibinfo {volume} {26}},\ \bibinfo
  {pages} {851} (\bibinfo {year} {1971})}\BibitemShut {NoStop}%
\end{thebibliography}
\end{document}